\documentclass[aps,prd,10pt,a4paper,altaffilletter,amssymb,showpacs,nofootinbib,twocolumn,showkeys,a4paper,floatfix]{revtex4-1}
\pdfoutput=1
\usepackage{amssymb,amsmath,amsfonts}
\usepackage{subfigure}
\usepackage[utf8]{inputenc}
\usepackage[english]{babel}
\usepackage{amsfonts}
\usepackage{amsbsy}
\usepackage{color}
\usepackage{multirow}
\usepackage{rotating}
\usepackage{slashbox}
\usepackage{ctable}
\usepackage{ulem}

\newcommand{\be}{\begin{equation}}
\newcommand{\ee}{\end{equation}}
\newcommand{\bea}{\begin{eqnarray}}
\newcommand{\eea}{\end{eqnarray}}

\newcommand{\ben}{\begin{eqnarray}}
\newcommand{\een}{\end{eqnarray}}

\begin{document}

\title{Revisiting a model-independent dark energy reconstruction method}

\date{\today}

\author{Ruth Lazkoz}
\author{Vincenzo Salzano}
\author{Irene Sendra}
\affiliation{Fisika Teorikoaren eta Zientziaren Historia Saila, Zientzia eta Teknologia Fakultatea, \\ Euskal
Herriko Unibertsitatea, 644 Posta Kutxatila, 48080 Bilbao, Spain}

\begin{abstract}
Model independent reconstructions of dark energy have received some attention. The approach that addresses the reconstruction of
the dimensionless coordinate distance and its two first derivatives using a polynomial fit in different redshift windows is well
developed \cite{DalyDjorgovski1,DalyDjorgovski2,DalyDjorgovski3}. In this work we offer new insights into the problem by focusing
on two types of observational probes: SNeIa and GRBs. Our results allow to highlight some of the intrinsic weaknesses of the method. One
of the directions we follow is to consider updated observational samples. Our results indicate than conclusions on the main dark energy
features as drawn from this method are intimately related to the features of the samples themselves (which are not quite ideal). This
is particularly true of GRBs, which manifest themselves as poor performers in this context. In contrast to original works, we
conclude  they cannot be used for cosmological purposes, and the state of the art does not allow to regard
them on the same quality basis as SNeIa.
The next direction we contribute to is the question of how the adjusting of some parameters (window width, overlap, selection criteria) affect the results. We find again there is a considerable sensitivity to these features. Then, we try to establish what is the current redshift range for which one can make solid predictions on dark energy evolution. Finally, we strengthen the former view that this model is modest in the
sense it provides only a picture of the  global trend. But, on the other hand, we believe it offers an interesting complement to other approaches
given that it works on minimal assumptions.
\end{abstract}



\maketitle{}
\section{Introduction}

It is very timely to continue to explore the acceleration of the Universe \cite{nobel}, as its discovery has very recently finally gathered the top recognition of the community.
The question of whether  such evolution is produced by a dark energy component or rather a manifestation of bold unknown Physics remains one of the most puzzling topics in Cosmology 
\cite{derev}. Assuming the dark energy scenario, and beyond theoretical questioning about its physical nature,
it is most vital to learn  to extract  the best and most relevant information from current astronomical data.
Many methods are strongly tied to the assumption of a background cosmological model or, equivalently, of
an analytical dark energy equation of state (EoS). Such is precisely
the case of the reconstruction of the Hubble diagram of type Ia Supernovae (SNeIa), which provided
the first evidence for the acceleration of the Universe  \cite{nobel}. The diagram presents the corresponding distance modulus at different redhsifts
and is computed from the EoS itself after a convenient integration in redshift.
Variations of this procedure apply as well when considering other observational tests
such as gamma ray bursts (GRBs), baryon acoustic oscillations (BAO) or the cosmic microwave background (CMB).
But even though biases on final results may depend on the previously described assumptions, that route, and
in particular giving a
specific prescription  of the EoS is still the most popular data analysis method, as opposed
to model independent methods (EoS free prescriptions) for the reconstruction of dark energy.
Fortunately, literature sources about this alternative way to explore the accelerated  universe
have mounted considerably, and thus the state of the art has turned promising \cite{Model}.

In \cite{DalyDjorgovski1,DalyDjorgovski2,DalyDjorgovski3} the authors present a different approach to this model independent reconstruction:
the dimensionless coordinate distance and its first and second order derivatives with
respect to redshift are reconstructed using a polynomial fit inside an appositely designed redshift window.
The only underlying assumption is  a description of the Universe by a generic
Friedmann-Robertson-Walker (FRW) metric.
Moreover, the method does not depend  on the physical properties of the  components of the Universe
or on the underlying gravity theory. The latter is an aspect to be considered
only at a second stage, when one attempts at the derivation of the behavior of more interesting and observable cosmological quantities:
the dimensionless expansion rate $E(z) \equiv H(z)/H_{0}$, the universe acceleration (through the well known
deceleration parameter $q(z)\equiv -{\ddot{a}a}/{\dot{a}^{2}}$), or finally the dark energy EoS parameter
$w(z) \equiv p_{de}/\rho_{de}$.  As usual, here $a$ is the scale factor, the dots denote
derivatives with respect to cosmological time, and $p_{de}$ and
$\rho_{de}$ are respectively the dark energy pressure and density.

In this work we aim at a thorough and novel  study of a number of main aspects of this problem. Firstly,  observational
samples are updated, and results with \cite{DalyDjorgovski3} are compared, as far as possible. This lets us  verify that, opposite to what one could
expect \textit{in principle},  conclusions in general are strongly related to the way  observational samples have been built on.
This is not, thus, a caveat of the reconstruction method, and could rather be useful at a first observational
level so as to discern the requirements a sample needs to be useful for the inference of information about dark energy.
Secondly, the validity of GRBs  in the context of this method is checked. It has been pointed out \cite{DalyDjorgovski3}
that these data can be a promising tool to study cosmology at high redshifts.
In this regard, our results seem to hint that GRBs are not as compelling as desirable, and  probably  the reasons
for this are intrinsic to the nature of actual GRBs observables. This is in agreement with the critical view in \cite{Schaefercrit}.

After this, it is argued that the analysis in the model independent spirit of
\cite{DalyDjorgovski3} requires adjusting some parameters. But typically, this causes noise and large errors in some regimes
(this seems to be intrinsically related to the procedure itself). Thus, the following questions arise quite naturally:
how can the real cosmological information be \textit{read off} from these noisy patterns? And how and when can the goodness and closeness to reality of the reconstructed profiles
be ascertained?

At the next stage,  we move on to explore the range of application of the method  and compare it with
the results obtained using a specific dark energy EoS. At the same time, we investigate how physically relevant
parameters  depend on the errors and noise of data. 

Finally, we conclude that this method just provides an idea of the global trend of the parameters.
The large correlation among them does not allow to obtain independent measures. But, on the other hand, it is admissible to report
values associated with independent redshift bins.

In Sec.~\ref{sec:Method} we revise the method in its main aspects; in Sec.~\ref{sec:Data} we briefly describe the  observational tools chosen for the analysis, namely the SNeIa and GRBs samples with their properties; in Sec.~\ref{sec:Results} we discuss the main results both for real and mock data; in Sec.~\ref{sec:Binning} we present results from the binning approach; and in Sec.~\ref{sec:Conclusions} we summarize the most important results.

\section{Method}
\label{sec:Method}

\subsection{Basic quantities}

The use of SNeIa and GRBs makes us focus mainly on
the distance modulus $\mu$, which is an observable quantity defined as
\begin{equation}\label{eq:modulus}
\mu(z) = 5 \log_{10} [ D_{L}(z)] + \mu_0 \; .
\end{equation}
Here $D_{L}$ is the luminosity distance (see below),
the nuisance parameter $\mu_{0}$ encodes the  absolute magnitude $M$ of any SNeIa
and the Hubble
constant $H_0$  (the current value of the expansion rate). For convenience, the speed of light has been set to one,
and this choice will apply to the whole paper.

In full rigour, though, a proper model-independent
analysis of the GRBs data cannot be based on the GRBs distance modulus, which
is not directly observable and  depends on the cosmology assumed \cite{GRB-No}.
As we will discuss in \ref{sec:GRBsdata}, we have used a GRBs sample whose calibration
is independent of the cosmological model and is based on the use of SNeIa as calibration tools.

The luminosity distance can be written as
\begin{equation}\label{eq:l_distance}
D_{L} = H_{0}^{-1}(1+z) \, y(z).
\end{equation}
where  the dimensionless coordinate distance $y(z)$ reads explicitly
\begin{equation}\label{eq:coordinate_distance}
y(z) = \int_{0}^{z} \frac{\mathrm{d} z'}{E(z')} \; .
\end{equation}
The determination of the coordinate distances from the underlying observational data
typically proceeds under the assumption of a cosmological model, which reflects in the analytical
expression of $E(z)$, the dimensionless expansion rate.

The method developed in \cite{DalyDjorgovski1} does not require any
assumption on the cosmological model, or on the expression of the dark energy functional.
It is possible though to determine $y(z)$ in a model independent way. Then, as  determining
the  Hubble function $H(z)$,
and deceleration parameter $q(z)$ only involves the computation of the derivatives of $y(z)$, those two quantities
of interest inherit the model independence.

The method was developed by \cite{DalyDjorgovski1} and follows a series of  steps which we describe in the next subsection.

\subsection{Inputs and fits}
\textbf{\textit{Step 1.}}
The input data needed are  the collection of $(z_{i} \, , \, y_{i} \, , \, \sigma^{2}_{i})$,
quantities, where the index $i$ ranges from $1$ to $\mathcal{N}$ (the number of objects in the chosen dataset).
Let us describe those quantities one by one. Clearly, the $z_{i}$ are the redshifts for the dataset objects. Then, the
$y_{i} = y(z_{i})$ denote  the coordinate distance data values obtained from the observational quantities (as the
        observed variable is $\mu(z)$, Eqs.~(\ref{eq:modulus})~-~(\ref{eq:l_distance}) have to be combined to give the $y_{i}$).
Finally, the $\sigma^{2}_{i}$ represent the coordinate distance errors. Contributions to this last quantities will come both
from errors on $\mu(z_{i})$ and from errors on the redshifts $z_{i}$. The sample we have used was not presented with errors on the redshifts, but using
        old dataset values, we have verified that the redshift contribution from those errors is always negligible
        when compared to the contribution from the errors in the $\mu(z_{i})$.

\textbf{\textit{Step 2.}} We define a redshift grid with the values $z_{j}$, and  at each of them we
        evaluate the coordinate distances, their derivatives and the related cosmological quantities. The properties
        of this grid are completely arbitrary, but we have checked this does not affect final results; in our case the grid will range from $0$
        until the maximum redshift $z_{max}$ reached by the chosen dataset, with an homogeneous spacing of $0.005$.

\textbf{\textit{Step 3.}} Finally, we have to perform a weighted polynomial fit of $(z_{i} \, , \, y_{i})$, with the inverse errors $1/\sigma^{2}_{i}$
        as statistical weights. The fit is not, though, a global one. We rather  do
  a number of individual fits in separate redshift windows defined by a length $\Delta_{f}$ and a center $z_{0}$. For each fit just those
  data points falling in each window are taken, and the central redshift moves iteratively  along the grid values $z_{j}$. At both ends of
        each  window, we apply in addition a Gaussian tapered region with dispersion $\sigma_g$ and extending out
        to $2 \sigma_g$. This makes all the points in this region have their weights reduced by the value of the Gaussian wing at that point. This way,
        fluctuations and discontinuities are controlled when moving from one window to another.
        Then the fit is performed after changing the variables from $z_{i}$ to $x_{i} = z_{i} - z_{0}$, so that the coordinate
        distance values and its derivatives at $z_{0} = z_{j}$ are simply the values of the coefficients of the polynomial evaluated in $x = 0$. For example, when using a second order polynomial:
        \begin{equation}
        y(x) = a_{0} + a_{1} x + a_{2} x^2,
        \end{equation}
        we will have
        \begin{equation}
        y(0) = a_{0}, \quad y'(0) = a_{1}, \quad y''(0) = a_{2};
        \end{equation}
        where the prime denotes differentiation with respect to $x$ (or, equivalently, to redshift, in this case). All the cosmological
        quantities depend on combinations of the respective $(y(0),y'(0),y''(0))$, and the errors can be easily evaluated by
        straightforward application of error propagation theory.

\subsection{Outputs and derived quantities}
The method described in the previous subsection
produces a set of $j$ polynomial-reconstructed dimensionless coordinate distance values
with their first and second order derivatives values (here $j$ is an integer  depending on the redshift grid range and spacing).
This allows giving the next step in the reconstruction.

Assuming an FRW metric and no contribution from spatial curvature,
one can easily obtain estimations for more cosmologically interesting quantities. One of them is the expansion rate,
which from Eq.~(\ref{eq:coordinate_distance}) is easily derivable to be
\begin{equation}
E(z) = (y')^{-1},
\end{equation}
where the prime indicates differentiation with respect to redshift.
Likewise, the deceleration is given by
\begin{equation}
q(z) = -\frac{\ddot{a}a}{\dot{a}^{2}} = -1 -(1+z)\frac{y''}{y'} \;.
\end{equation}
This expression is obviously valid for any cosmological scenario based on an homogeneous
and isotropic expansion.
If we now also assume a theory of gravity (in this case General Relativity), we can use the Friedmann equations:
\begin{equation}
\frac{\ddot{a}}{a} = -\frac{4 \pi G}{3} (\rho_{m} + \rho_{DE} + 3 p_{DE}) \, ,
\end{equation}
\begin{equation}
\left(\frac{\dot{a}}{a}\right) = \frac{8 \pi G}{3} (\rho_{m} + \rho_{DE}) \, ,
\end{equation}
where $\rho_{m}$ is the mass-energy density of non-relativistic matter, $\rho_{DE}$
is the mass-energy density of the dark energy, and $p_{DE}$ is the pressure of dark energy.
It is straightforward to derive from them the expression of the dark energy equation of state $w$. Using
the standard definition of the critical density at the present epoch, $\rho_{0,c} = 3 H_{0}^2 / (8 \pi G)$,
 the dark energy pressure can be shown to be
\begin{eqnarray}
p_{DE}(z) &=& \rho_{0,c} \left[ \frac{E^{2}(z)}{3} (2 q(z) -1) \nonumber \right]\\
&=& \rho_{0,c} \left[ - (y')^{-2} \left( 1+ \frac{2}{3} (1+z) \frac{y''}{y'} \right) \right]\; ;
\end{eqnarray}
while the dark energy density is
\begin{equation}
\rho_{DE}(z) = \rho_{0,c} \left[ (y')^{-2} - \Omega_{m}(1+z)^3 \right] \; ,
\end{equation}
where $\Omega_m=\rho_{m}/\rho_{0,c}$ is the fractional contribution of non-relativistic
matter to the critical density at zero redshift, and the term $(1+z)^3$ defines its time
evolution. Then, as the dark energy EoS parameter $w(z)$ is defined as the ratio of the dark energy
pressure to density, $p_{DE}(z)/\rho_{DE}(z)$, it finally results in the expression
\begin{equation}
w(z) = \frac{-\left[1+(2/3)(1+z)y''(y')^{-1}\right]}{1-\Omega_m(1+z)^{3}(y')^{2}} \; .
\end{equation}
Note that  $\Omega_m$  is clearly involved in the determination of the $w(z)$ parameter and of its related error; here
we use the value corresponding to the wcdm+sz+lens case from WMAP7-year
\cite{wcdm}, which has $\Omega_m= 0.259^{+0.099}_{-0.095}$. Nevertheless, this particular choice
does not have a strong impact in the conclusions to be reached.

To close this section let us mention that errors on all the previous quantities are of course derived from errors from $y$, $y'$
and $y''$ applying the usual error propagation rules.

\section{Current Data}
\label{sec:Data}

\subsection{Supernovae}
\label{sec:SNdata}

Our main reconstruction tool is one of the most recent type Ia supernovae sample available, the Union2
sample described in \cite{Amanullah10}. This
compilation is the result of a new dataset of low-redshift nearby-Hubble-flow
SNeIa and is built with new analysis procedures suitable for working with several
heterogeneous SNeIa compilations. It includes the Union data set
from \cite{Kowalski08} with six added SNeIa first presented in
\cite{Amanullah10}, along with SNeIa from \cite{Amanullah08}, the low-z
and the intermediate-z data from \cite{Hicken09a} and
\cite{Holtzman08} respectively. After the application of various selection cuts
to create a homogeneous and high signal-to-noise
data set, $\mathcal{N}_{\mathrm{SNeIa}}=557$ SNeIa events distributed over the
redshift interval $0.015 \leq z \leq 1.4$ were obtained.

The use of the Union2 SNeIa data rests on the definition of the distance modulus as given above.
Given the heterogeneous origin of this dataset, and the procedures described in \cite{Kowalski08} for
reducing data, it turns out we need to fix the value of $\mu_0$ for converting the modulus
distance $\mu(z, \mu_0)$ to the coordinate distance $y(z)$. One way would be using just
low redshift SNeIa  to find the best fit value for $H_0$ starting from the approximate expression valid for this range,
$D_{L} \approx z/H_0$; alternatively, one could assume an independent
measurement of the same quantity, as in \cite{Riess:2011yx}.

Nevertheless, we have chosen a different way to proceed: as we will try to compare results from SNeIa with results
from GRBs, choosing the same unique value for $H_0$ makes sense, even though
it is not fundamental for the final results. For this reason, as the GRBs calibration
has been done assuming a well fixed value of $H_0 = 70$ (km/s)/Mpc \cite{Schaefer07,DalyDjorgovski3},
we have decided to use the same value of $H_0$ for the SNeIa  (which at the same time we have tested to give a very good
fit to the Hubble diagram of SNeIa at low redshifts).

\subsection{Gamma Ray Bursts}
\label{sec:GRBsdata}

Working out and interpreting results from a GRBs analysis is not an
easy task, as the errors on their observable quantities are much
larger than those for SNeIa, besides the fact their source mechanism
is still not well understood. For this reason, choosing a good GRBs
sample is crucial; we have chosen to work with the sample
described in \cite{Cardone09}. There the authors performed a new
calibration procedure on the widely used GRBs sample from
\cite{Schaefer07}, which matches perfectly some of the requirements in this paper.
The possibility of a comparison between
SNeIa and GRBs is strictly related to the building of a Hubble
diagram for GRBs too, which is an extremely difficult task because
GRBs are not standard candles, unlike SNeIa. To create an Hubble
diagram for GRBs, one has to look for a correlation between a
distance dependent quantity and a directly observable property.
Starting from some of the many correlations that have been
suggested not long ago, in \cite{Schaefer07} the authors created a sample of
69 GRBs in the redshift range $0.17<z<6.6$ for which the Hubble diagram is
well settled.

Then, in \cite{Cardone09} that sample was updated in many
aspects, the main one being the test of a new method for the
calibration of GRBs based on the non  assumption of  a
priori cosmological model. Such a model independent calibration
is built on the idea that SNeIa and GRBs at the same redshift
should exhibit the same distance modulus. In this way,
interpolating the SNeIa Hubble diagram gives the value of $\mu(z)$
for a sub-sample of GRBs which lies in the same redshift range.
This sub-sample can be finally used to calibrate the well known
GRBs correlations and, assuming that this calibration is redshift
independent, it can be extended to high redshift GRBs. With this
procedure, in \cite{Cardone09} the authors were able to convert the
\cite{Schaefer07} sample to a new one, with the same number of
objects but with SNeIa-calibrated GRBs distance moduluss.

\section{Results}
\label{sec:Results}

We remind that the main objective of this kind of approach is not to obtain
independent measurements of coordinate distances and their related
quantities, but rather to find out the global trends underlying
the data. More exactly, we aim to specify a possible way of proceeding for
optimizing the use of this approach. In particular, we will examine the effects
produced by changes in the fitting windows which should
enable to detect and distinguish \textit{global} and \textit{local} trends.

On the one hand, as it is well known, using larger fitting windows leads to more
robust fits, as more points are included in the analysis; but, at the same time,
a strong correlation arises among the various values one obtains, as the
fitting windows clearly overlap.

On the other hand, using narrower fitting windows, makes it possible
to find local (in a redshift sense) properties of the global trend,
and with more redshift resolution. The negative counterpart
is that smaller windows also imply working with fewer data points,
and this accentuates noisy features in the estimated parameters.

If one assumes implicitly the physically reasonable hypothesis
that the coordinate distance $y(z)$ is a smooth function of redshift (as one
does in this kind of approach), then one can compare large (global) and small (local)
fitting windows results. In that way, one can explore the global trend and the
characteristics that depend on local irregularities coming from, for example, mixing
heterogeneous sub-samples, even if they are treated appropriately, as is the case of the Union2 sample,
or containing intrinsic and physically important information about the universe
dynamics. By comparing results from large and small fitting windows, one can detect more clearly reconstruction features
due to the limitations of the method rather than to an intrinsical relation to the physics dynamics.
In this work we study which is the best range to obtain this information, and we also
gather some further understanding about the limits of the
method presented: we basically conclude the limits
depend on the characteristics of the datasets and  the specifications of the fitting windows.
We have also performed an independent binning sketch, which has consisted in dividing
the data points into a number of bins which do not overlap, and
obtaining  independent estimations of the required parameters afterwards.

\subsection{SNeIa: comparison with literature}

To avoid misunderstandings, let it be clear that there are a number
of facts that preclude a very close  comparison of our results with those in \cite{DalyDjorgovski3}.
To begin with, our SNeIa sample has two main differences with respect to the
samples used in \cite{DalyDjorgovski3}. First of all, the sample we consider is made up of $557$ objects as opposed
to the $192$ in  Davis' dataset. This mainly implies a higher mean number density
for the data points which should translate into a better knowledge of the global tendency of the parameters under study. This newer
sample displays a quite larger number density for $z<1$, but no considerable changes occur for higher redshifts values,
so one still has to be careful about conclusions based on the results in this range.
Second, the sample we use is made of different heterogeneous SNeIa sub-samples and, even though data reduction
has reduced differences considerably, it is possible that
those disparities influence the local properties of our derived quantities.

A quite important new point in our analysis is that we have noticed that some remarkable oscillations that arise
at some levels of the reconstruction are not
directly due to the fitting algorithm, but mainly related to the presence of SNeIa with a high relative error.
To back up this view, we produced a cut version of the total Union2 data sample by deleting
all the data points whose error was larger than $<err> + \; 2 \; \sigma_{err}$, where $<err>=0.225$ is
the mean of the data error distribution and $\sigma_{err}=0.133$ is its standard deviation. With this procedure
we cut $30$ points (all having in common a high relative error) and the oscillatory (noisy) features disappeared.

We performed different runs changing both the length of the fitting windows, $\Delta_{f}$, and the dispersion
$\sigma_g$ of the Gaussian tapered region needed for calculating the fit weights.  We regarded as  good results just those
obtained when the number of data-points inside any fitting window and out of the Gaussian tapered region
(i.e. with unchanged error weight) was $\geq 10$, so to assure a
minimal sufficient number of data points as compared to the number of fitting parameters. Our choice is slightly stricter than
the one proposed in \cite{DalyDjorgovski1}.

\newcommand{\figjump}{\addtocounter{\thisfig}{1}(\alph{\thisfig})\\\\}
\newcommand{\thisfig}{fig:DDcomparison}
\newcounter{\thisfig}
\begin{figure}
\begin{tabular}{c}
\includegraphics[width=7.5cm]{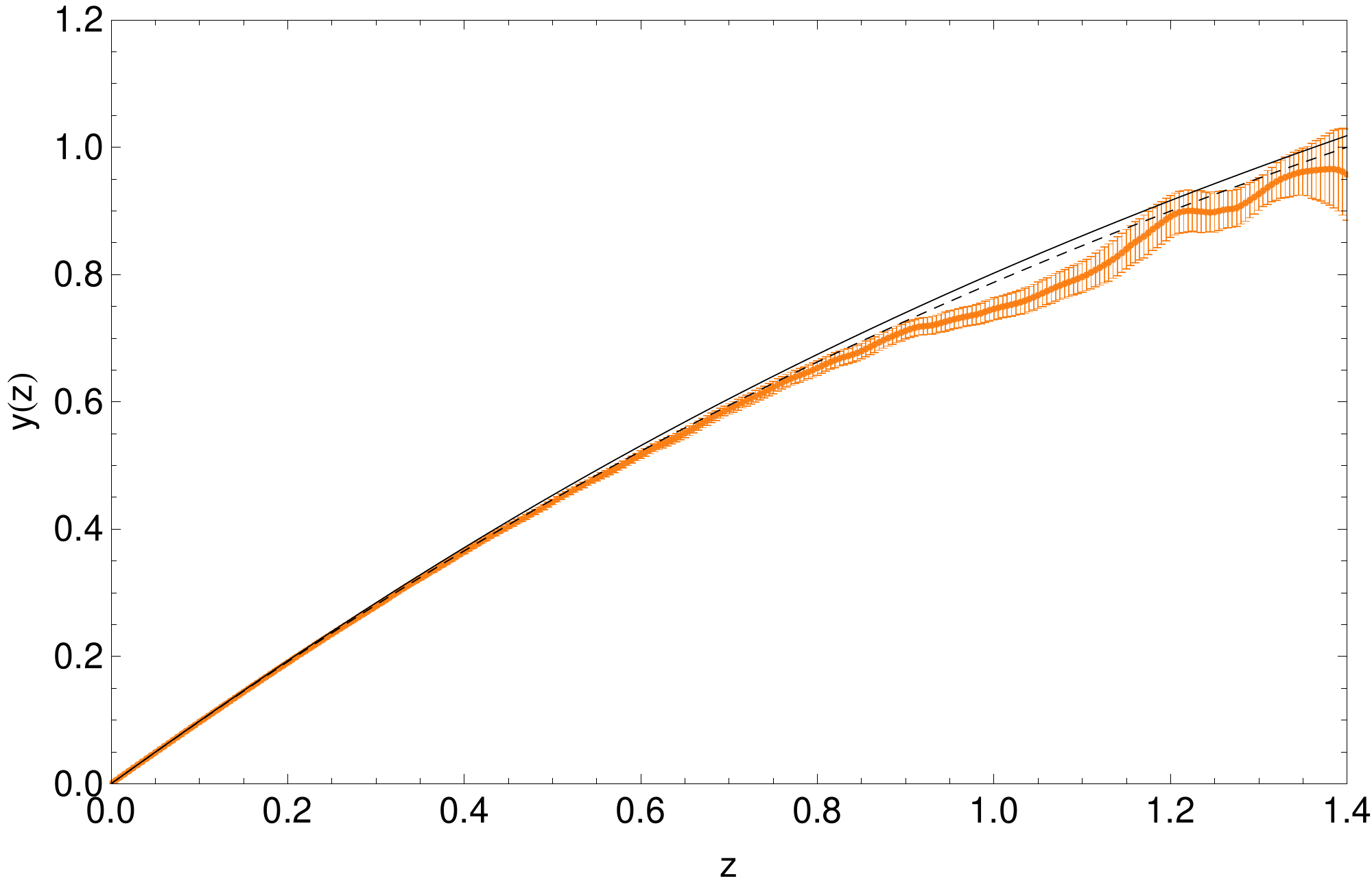}\\
\figjump
\includegraphics[width=7.5cm]{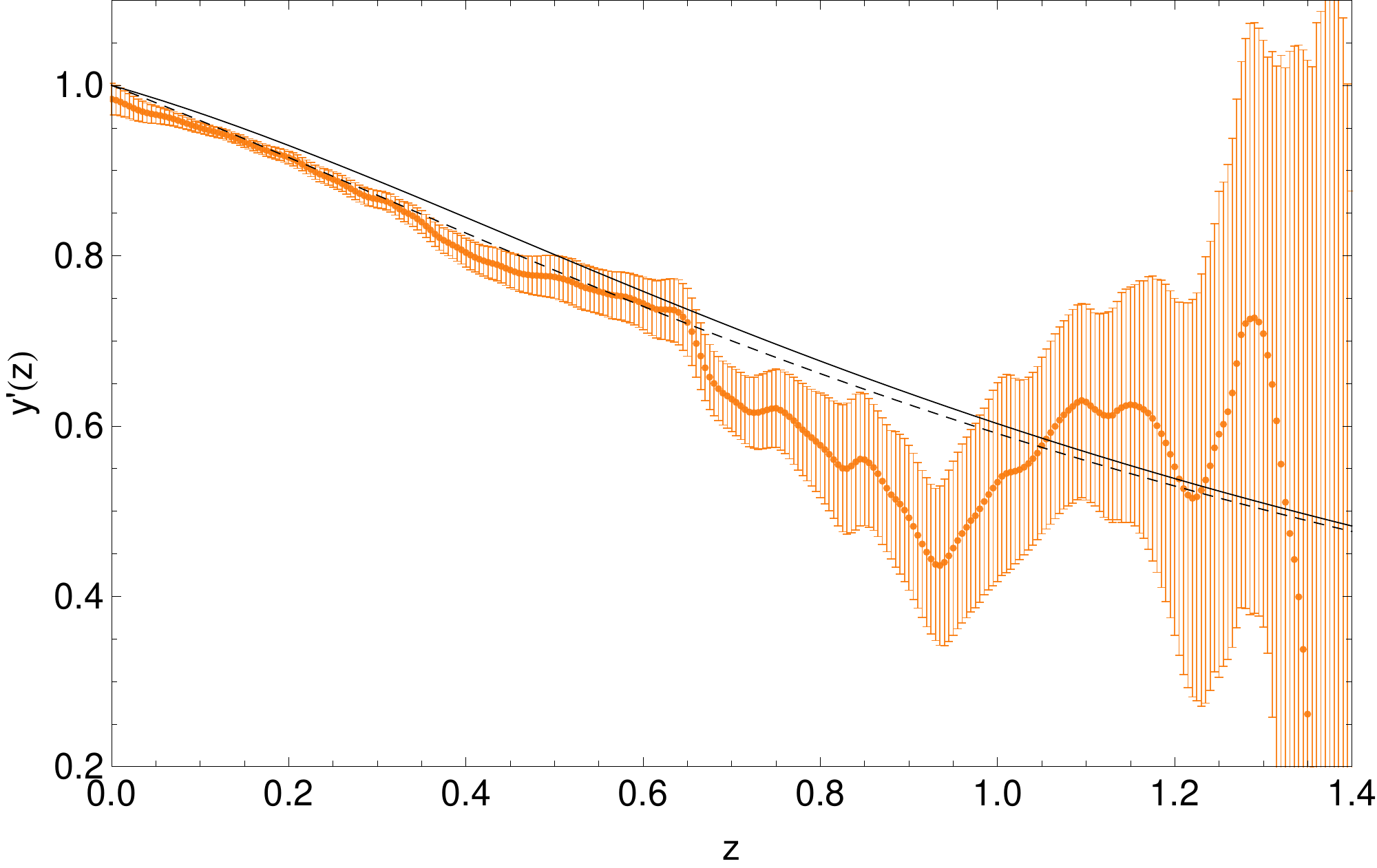}\\
\figjump
\includegraphics[width=7.5cm]{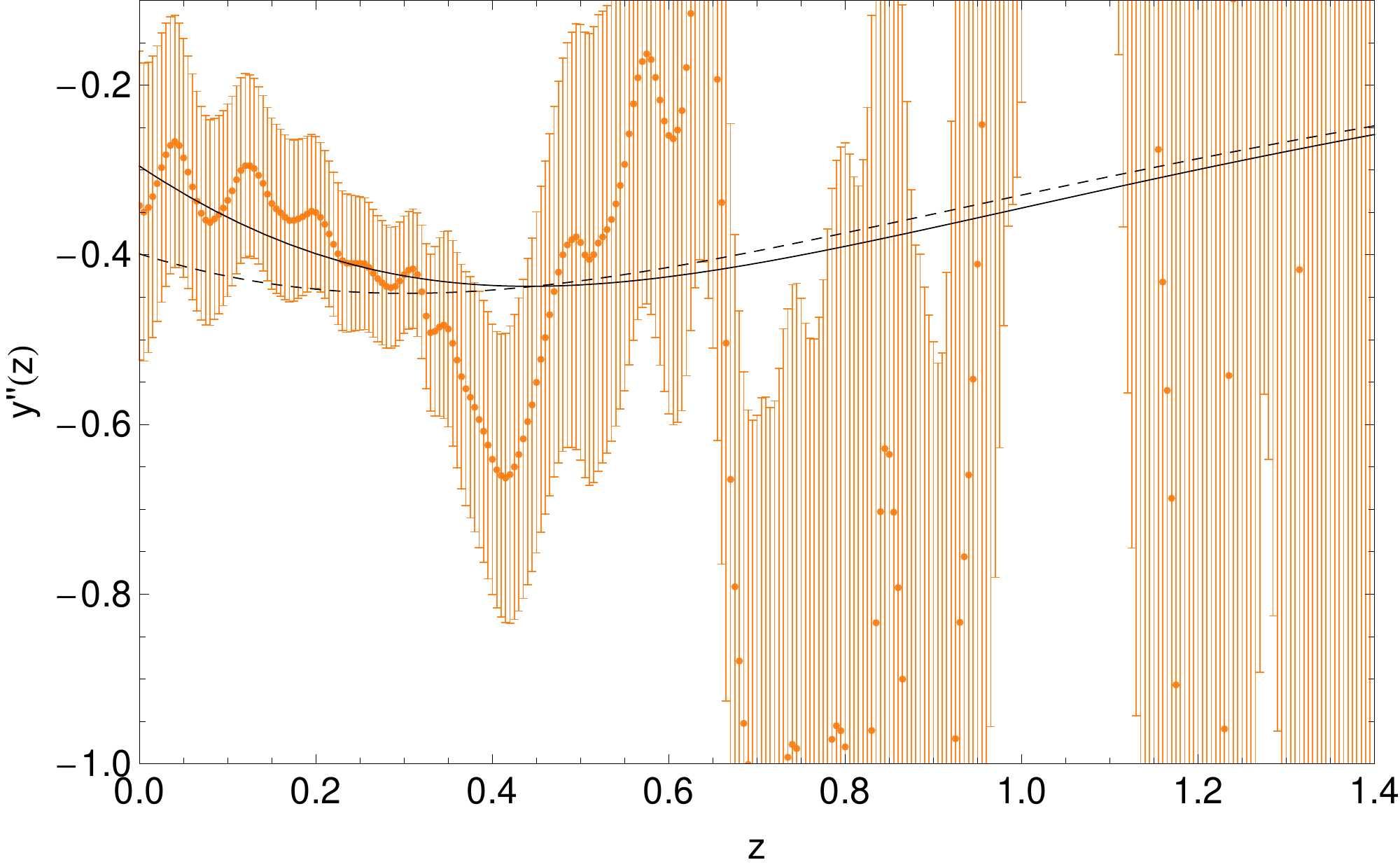}\\
\figjump
\end{tabular}
\caption{Comparison of $y(z)$, $y'(z)$ and $y''(z)$ with $\Delta_{f}=0.6$ and $\sigma_g=0.02$ from a polynomial analysis with a WMAP7 $\Lambda$CDM ($\Omega_m = 0.266$, solid line) and with a WMAP7 ($+$ WMAP7.2) quiessence model ($\Omega_m = 0.27$, $w = -1.1$, dashed line).}
\label{\thisfig}
\end{figure}

Fig.~\ref{fig:DDcomparison}(a) shows a reconstruction of $y$ derived from a fit with $\Delta_{f} = 0.6$
and a dispersion $\sigma_g = 0.02$ of the Gaussian tapered region (i.e. the same values used in \cite{DalyDjorgovski3} and regarded there as the
\textit{best} ones). The trend is clearly very smooth and quite similar to the one in \cite{DalyDjorgovski3}; all in
all, it can be considered as an admirably good reconstruction.
We detect, though, that for redshifts $z>0.8$ there is a slight difference between the reconstructed dimensionless coordinate
distance and the theoretically expected one (derived using some updated cosmological models from \cite{wcdm} with respect
to the one used in \cite{DalyDjorgovski3}), with a systematic lower trend at high redshifts. Given the location (redshift range) of
the deviation, we believe it can put be down to the low number density of SNeIa points in this region.

The reconstruction of the first order derivative of dimensionless coordinate distance $y'$ is clearly less satisfactory.
We plot it in Fig.~\ref{fig:DDcomparison}(b) using the same redshift range as for the corresponding figure in \cite{DalyDjorgovski3}.
Both studies display in a broad sense similar irregularities for $z>0.6$, but a rise
at high redshifts occurs in our case;
it is either a hint of  a possible systematics
in the way high redshift data have been obtained, or it is rather a natural consequence of the low number of data
points per bin used in the interpolation, or just a mixture of those such problems. No way of disentangle the two effects is possible at this level.

If we compare this trend in $y'$
with the results about $y$, it follows immediately that the redshift
range where $y'$ rises is the same where $y$ deviates more from the theoretical expectations. We underline that the
theoretical expectations (plotted in the same figures) are only considered as a possible comparison term and a tool for detecting possible
local or global properties of the underlying trend. Their use does not mean they are considered important
as a fiducial cosmological model.

For what concerns the second order derivative, $y''$, Fig.~\ref{fig:DDcomparison}(c),
we clearly observe a very noisy behavior at redshift $z>0.4$. This quantity seems more sensitive
to oscillations in the number density in SNeIa sample. We reach this conclusion because while
 \cite{DalyDjorgovski3} this  trend looks quite homogeneous, in our reconstruction it is more accentuated
so we are tempted to put it down to  the heterogeneous nature of the Union2 sample. Thus, even though we are working
with a larger SNeIa sample,  error bars are quite comparable
and, in our case, even larger at high redshift. So, the first lesson one can derive is that denser and larger
(in a number data point sense) samples might not improve the use of the model independent reconstruction method we are
analyzing. This can be seen as a limitation of the method itself, and in \ref{sec:Mock} we look deeper into this problem.

\subsection{SNeIa: general overview}

Continuing with our analysis, we have also considered differences arising  from the use of different  window width values.
After several preliminary tests,  we have eventually considered as representative results from
$\Delta_{f} = 0.6$ (which is the width
used in \cite{DalyDjorgovski3}), and from  another two widths: $\Delta_{f} = 0.4$ and $\Delta_{f} = 1.0$.

\subsubsection{How small a window?}
As we are working with a data set that contains a statistically significant number of objects,
one may be tempted to try and improve the analysis by using smaller redshift windows for the
fit, as, \textit{in principle}, one would still have a sufficient number of points.
This is, however, not true mainly due to the non uniformity of redshift distribution of our
SNeIa sample. We have a large number of objects at small redshifts but too few points are at high redshifts, and this makes
the noisy trends and the errors grow in this range. It is thus not surprising that  more stringent constraints
on our derived quantities at large redshifts cannot be obtained.

\begin{figure*}
\begin{tabular}{cc}
\includegraphics[width=7cm]{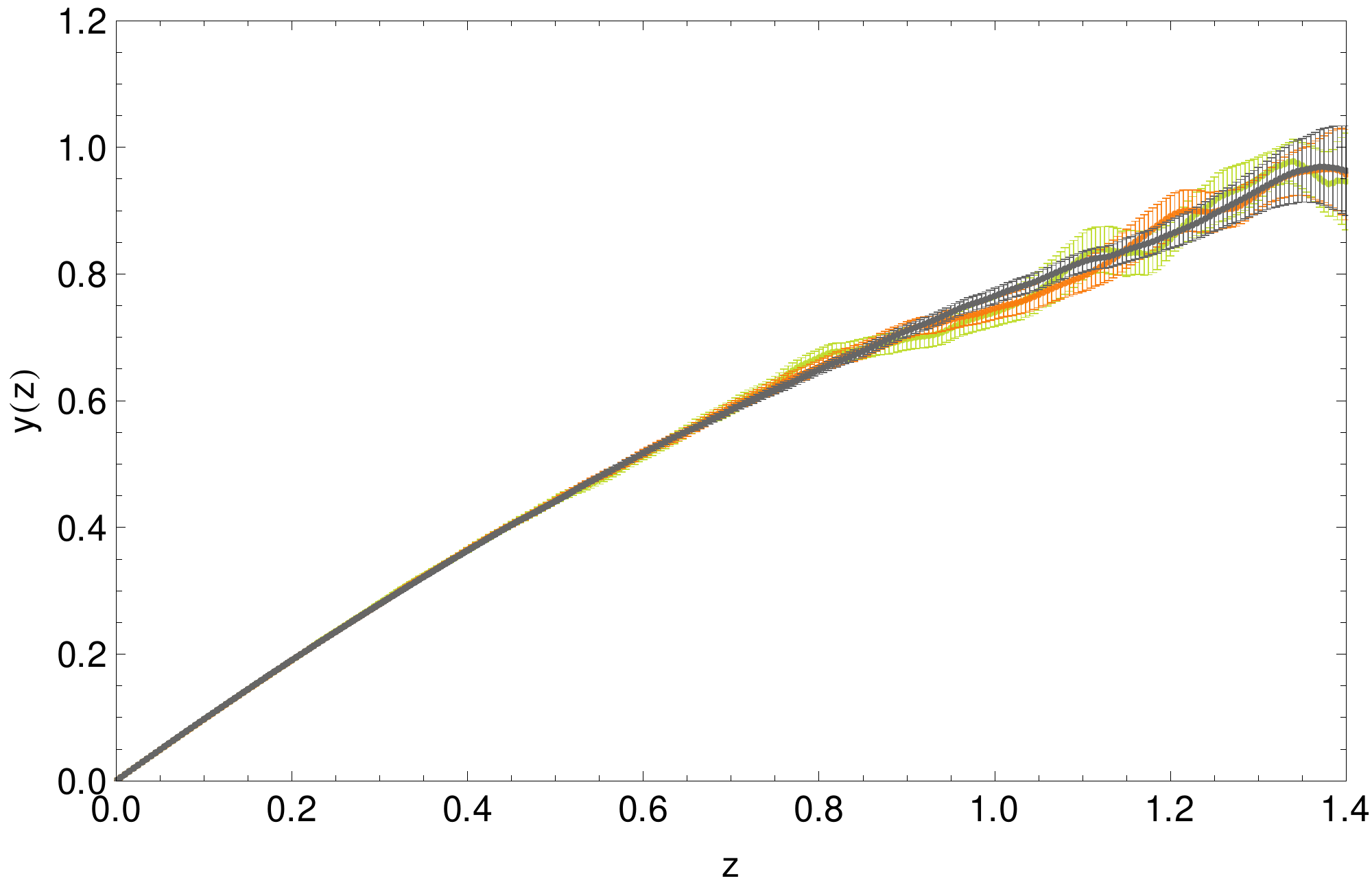}&
\includegraphics[width=7cm]{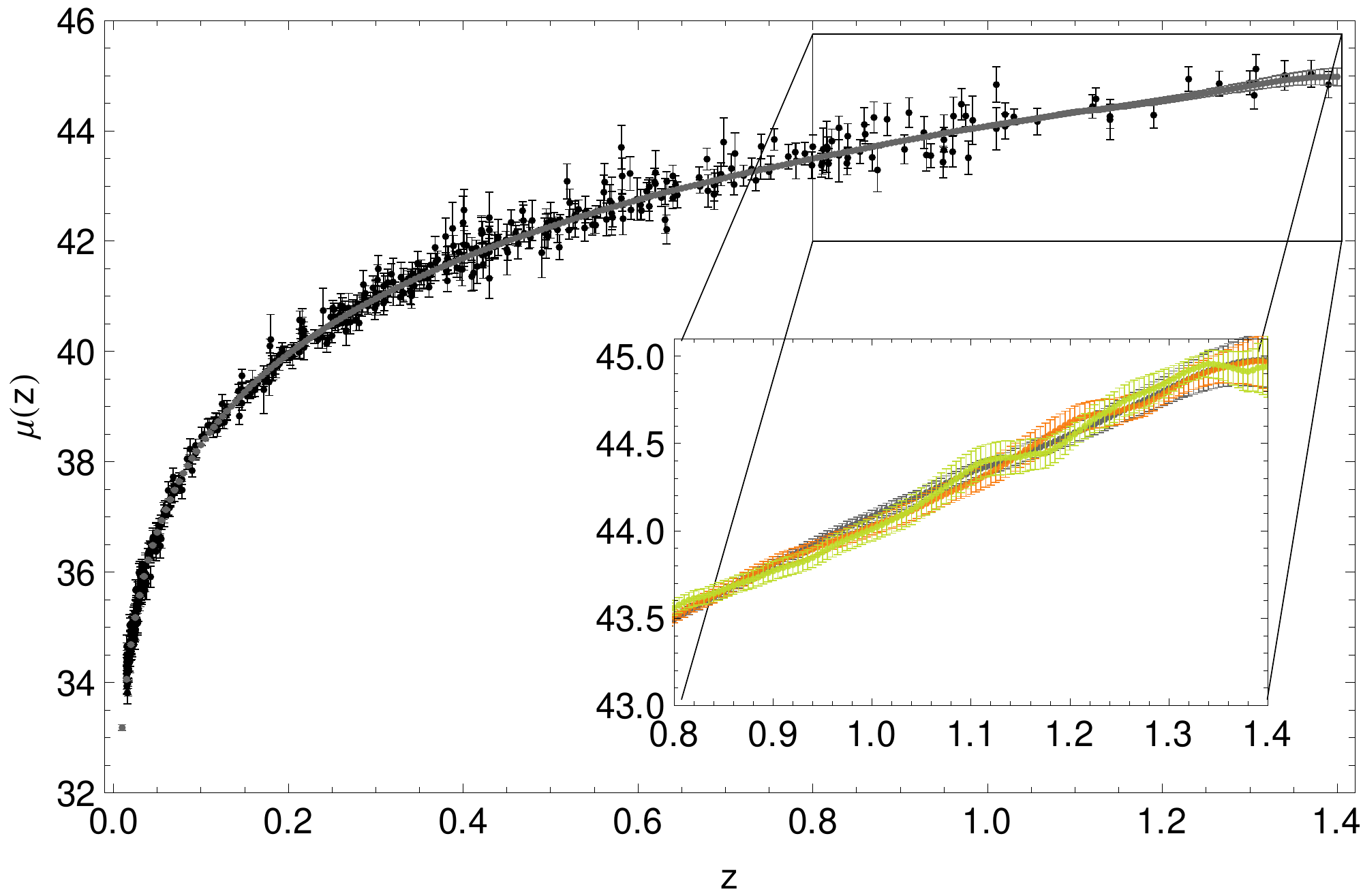}\\
(a)&(b)\\
\includegraphics[width=7cm]{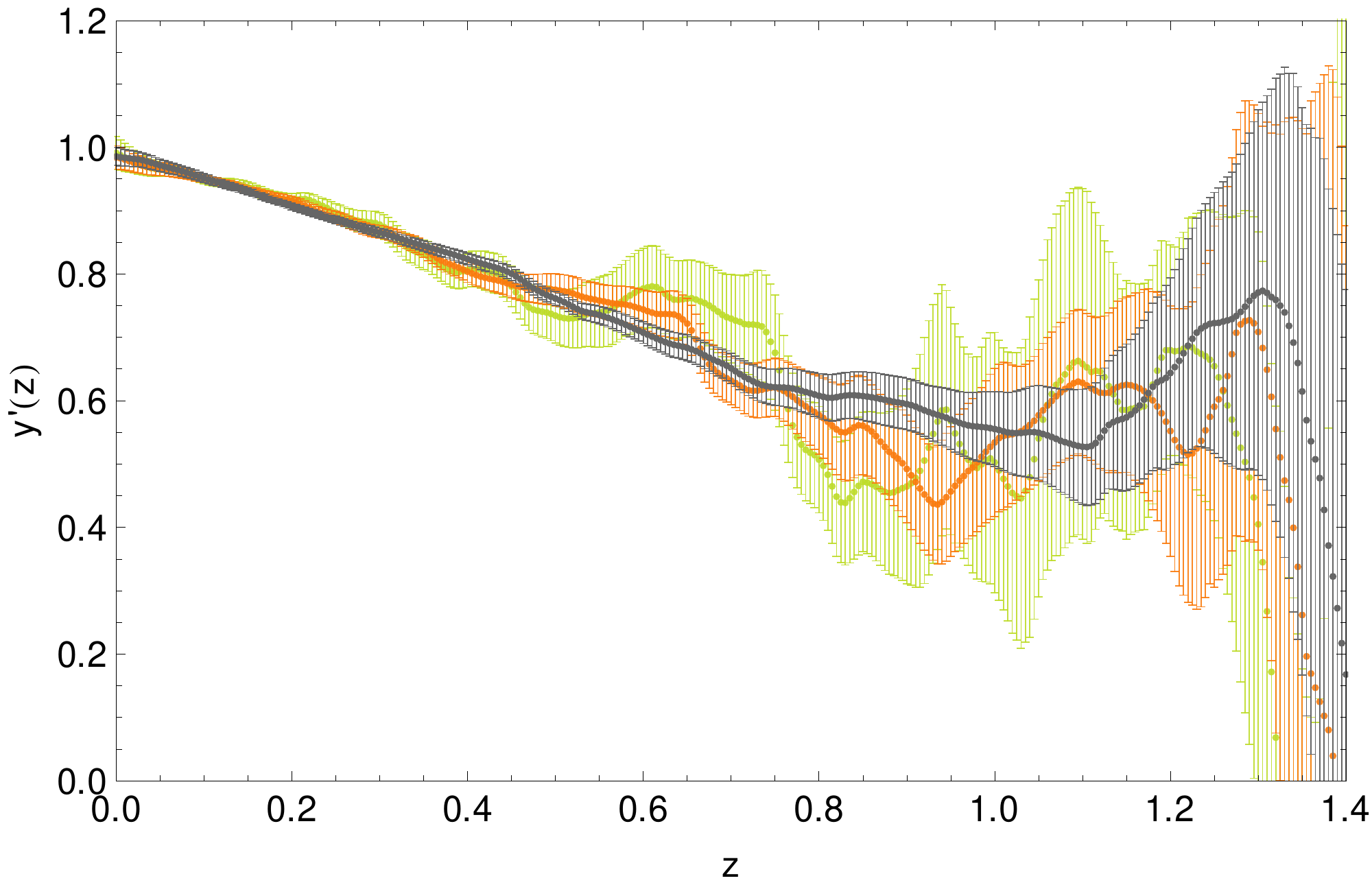}&
\includegraphics[width=7cm]{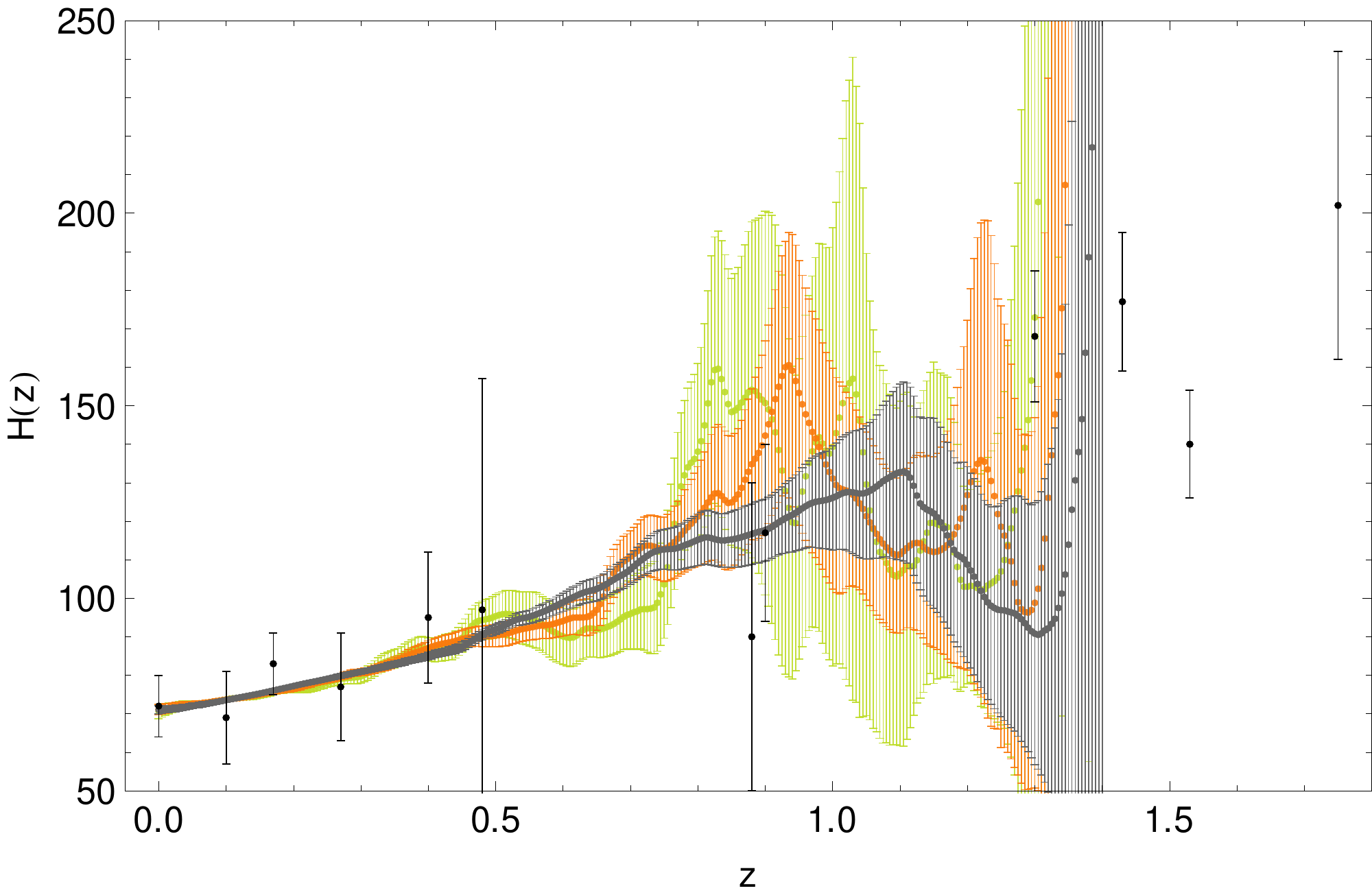}\\
(c)&(d)\\
\includegraphics[width=7cm]{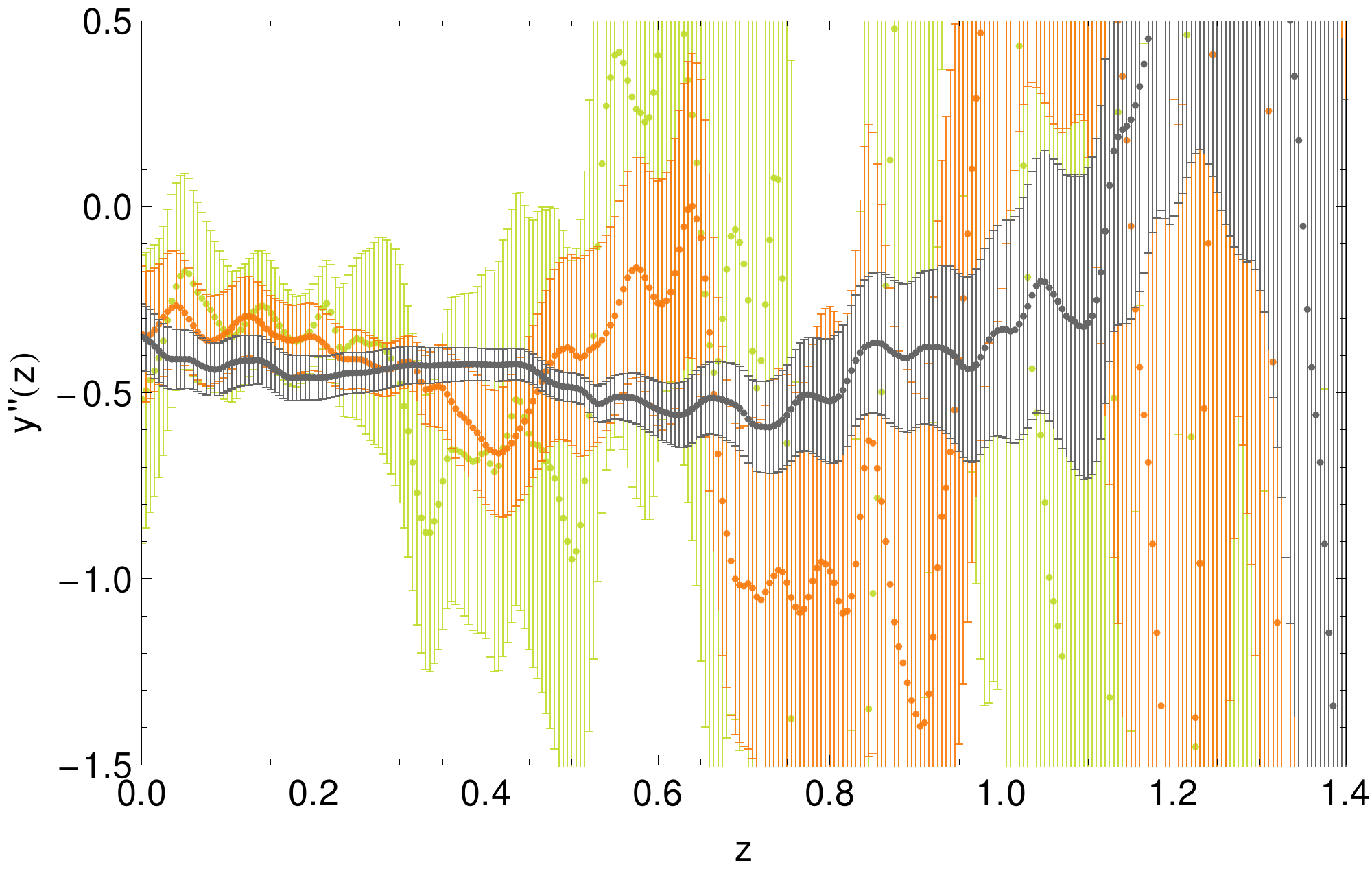}&
\includegraphics[width=7cm]{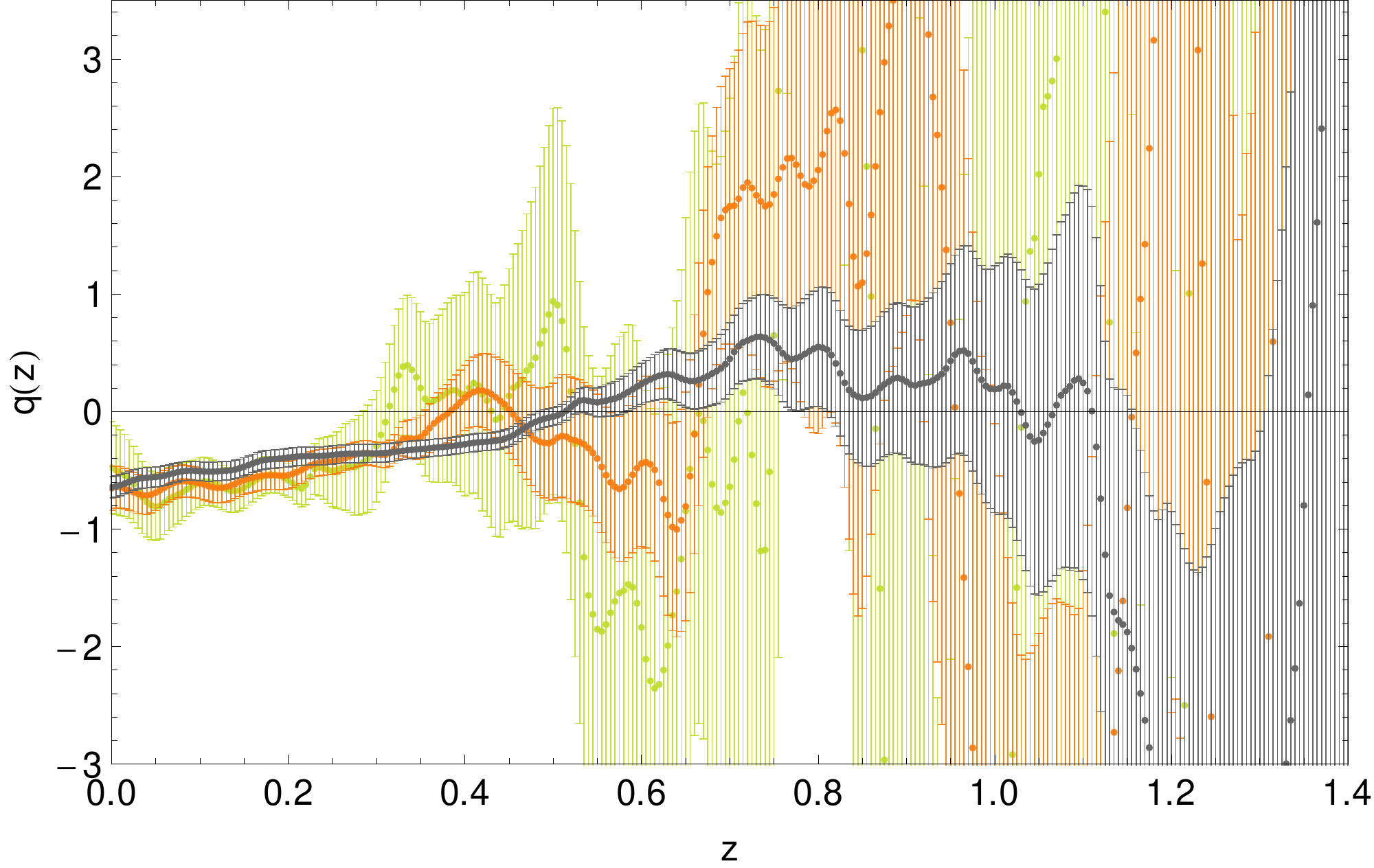}\\
(e)&(f)\\
\end{tabular}
\includegraphics[width=7cm]{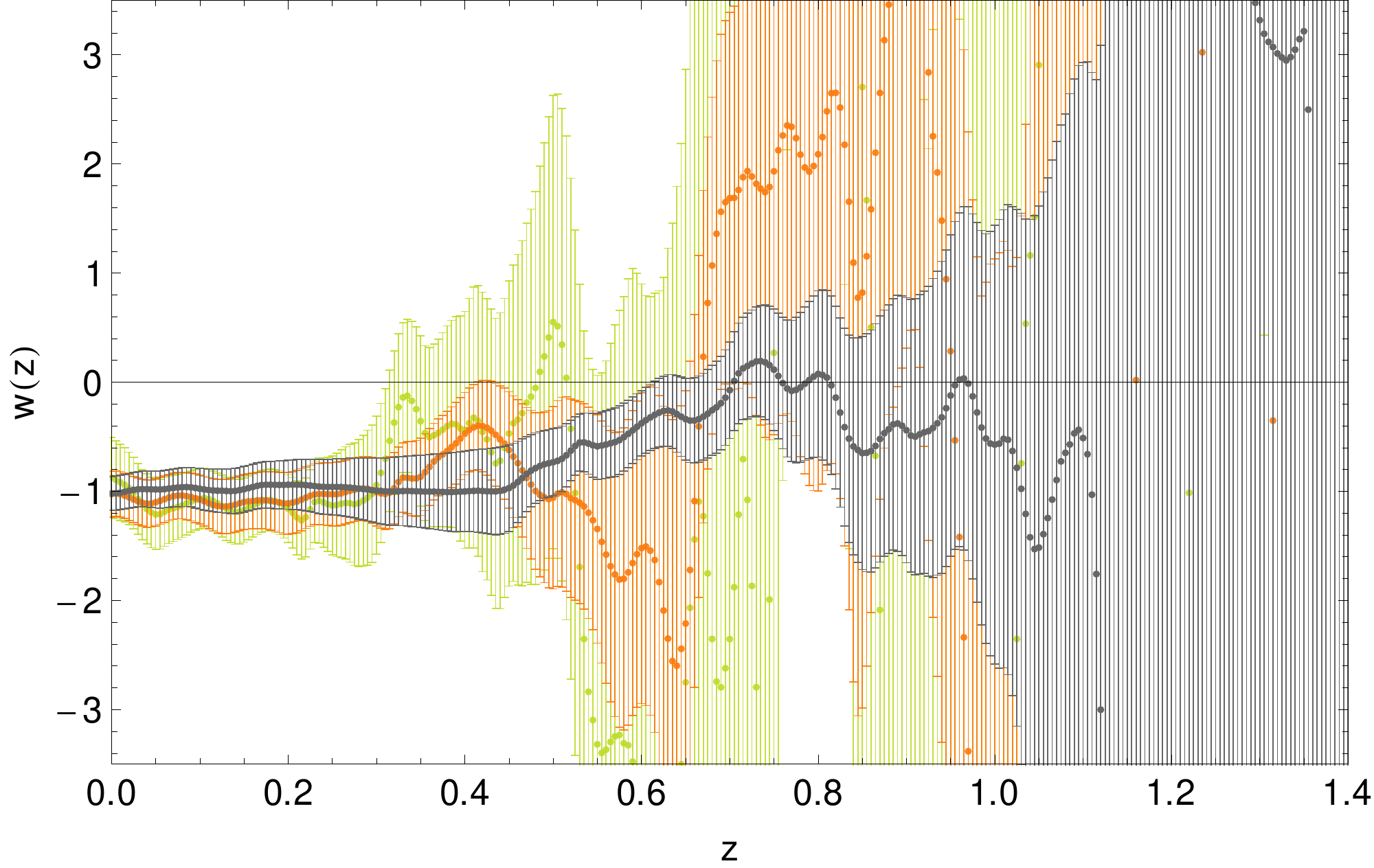}\\
(g)
\caption{Second order polynomial reconstruction with SNeIa (Union2). Coordinate distance $y(z)$ and its derivatives ($y'(z)$ and $y''(z)$), and the derived cosmological quantities: distance modulus $\mu(z)$,
Hubble function $H(z)$, deceleration parameter $q(z)$ and the dark energy EoS $w(z)$. Different colors are used for different fitting window $\Delta_{f}$: from lightest to darkest for $\Delta_{f}=0.4$, $\Delta_{f}=0.6$ and $\Delta_{f}=1.0$ respectively.}\label{fig:parameters2P}
\end{figure*}

Moreover, when the fitting window width becomes smaller, the \textit{extremal} intervals
(i.e. intervals at very low and high redshift) become precisely the most problematic cases.
We can see this in Fig.~\ref{fig:parameters2P}(e) by paying attention to the case of the reconstructed $y''$.
Note that the errors get minimal in the mid redshift range, but then grow again for low (large number of data points) and
high (small number of data points) redshifts.
On broad terms, at low redshifts the number
of datapoints in typical windows is high, and so we have small noisy oscillations and biased estimations of the reconstructed parameters, but
estimations also depend on the Gaussian tapered region dispersion, $\sigma_{s}$. In contrast
at high redshifts, few points will be found in our window, and constraints  are poor: many oscillations and large error bars occur.

\subsubsection{How large a window?}
On the other hand, taking into account that our main goal is to guess the global trend of our parameters, we
considered a  somehow completely opposite approach. Specifically, we studied the effect of enlarging
the fitting window as much as possible so as to detect a global and unified
trend of the data (even though this would cause overlaps in the  fitting windows and a related growth in the correlation).
We detected that large fitting windows give very smooth trends,
mainly because the data are, in some measure \textit{averaged}. But at the same time these results are strongly
biased (as already reported by   \cite{DalyDjorgovski3}),
because the polynomial fit describes  very poorly the general trend of our quantities. More specifically,
a second or third order polynomial can fit \textit{very well} the dimensionless coordinate distance in the redshift
range depicted by data, but it would give strongly unphysical trends when extended to higher (out of sample) redshifts
as, for example, a negative coordinate distance.
For this reason, we do not present here cases with widths  wider  that $\Delta_{f} = 1.0$,
because this is the largest amplitude for which such a bias has a negligible effect on our reconstructed quantities.

\subsubsection{Parameter estimates}

Let us review now our estimations of several parameters, obtained under the considerations described above.
\setcounter{secnumdepth}{5}

\paragraph{Hubble parameter}
\vspace*{0.25cm}
The reconstructed Hubble diagram hardly changes for
the three different windows lengths considered. As we show in the zoomed vision
of Fig.~\ref{fig:parameters2P}(b), 
differences are negligible and no significant discrimination is possible when compared with Union2. We have to recall
the good behaviour of the reconstruction of this quantity.
The reconstructed Hubble parameter, Fig.~\ref{fig:parameters2P}(d), is practically independent of the fitting
window for $z<0.5$, and in this redshift range it fits quite well the observational data
extracted from differential ages of passively evolving galaxies and discussed in
\cite{Stern10}. The Hubble constant (i.e. $H_{0} = H(z=0)$) is found to be:
\begin{eqnarray}H_{0} =\begin{cases}
     70.63 \pm 1.85 & \mathrm{for~~} \Delta_{f} = 0.4,\\
      71.14 \pm 1.34 & \mathrm{for~~} \Delta_{f} = 0.6,  \\
      71.04 \pm 1.02 & \mathrm{for~~}  \Delta_{f} = 1.0\;.
      \end{cases}~~~
      \end{eqnarray}
in units of (km/s)/Mpc. Clearly, differences among the three fitting windows are very small, just notice
that the larger the window, the lower the error on $H_0$.
We can also see that the three trends are quite similar up to $z \approx 0.5 $, while significant deviations appear for to high redshift data.

\begin{figure}[ht]
\centering
\includegraphics[width=8.5cm]{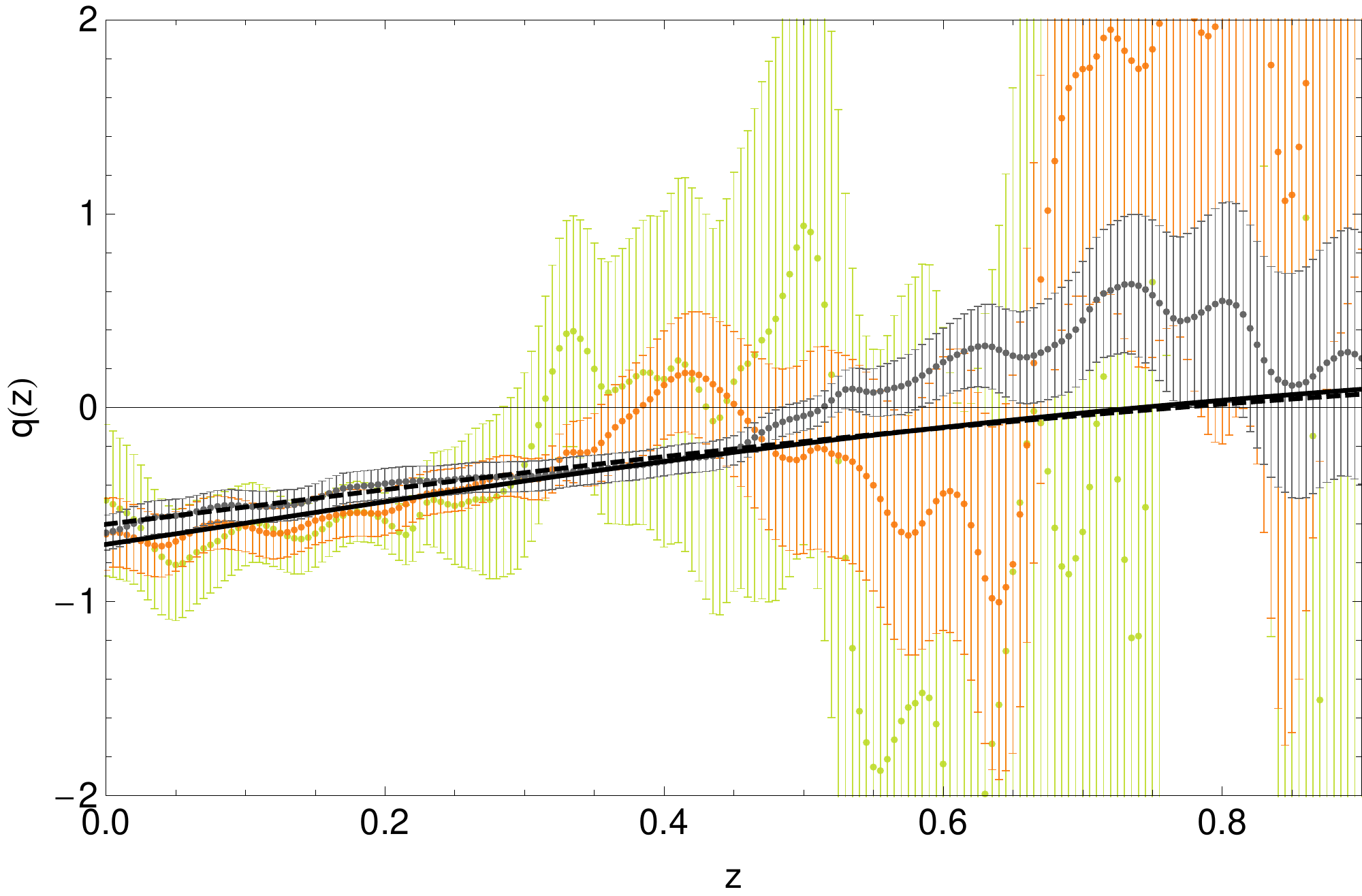}
\caption{Second order polynomial reconstruction with SNeIa (Union2): reconstruction of the deceleration parameter $q(z)$. Different colors are used for different fitting window $\Delta_{f}$: from lightest to darkest for $\Delta_{f}=0.4$, $\Delta_{f}=0.6$ and $\Delta_{f}=1.0$ respectively. The solid and the dashed black lines are
respectively the WMAP7 $\Lambda$CDM ($\Omega_m = 0.266$) and the WMAP7 ($+$ WMAP7.2) quiessence model.}\label{fig:Qwmap}

\end{figure}

\paragraph{Deceleration parameter}
The deceleration parameter $q(z)$, Fig.~\ref{fig:parameters2P}(f) is influenced by the noisy features in $y''(z)$, which become rather important at high redshifts.
Differences among the three  fitting windows considered become manifest when one calculates the present value of $q(z)$ and  $z_{t}$
(transition redshift from deceleration to acceleration\footnote{Given the noisy behavior it would be more correct to speak about a \textit{first-time} transition
redshift, as it is possible that such line is crossed many times more because of the large uncertainties than a real
physical source}. Our estimates are:
\begin{eqnarray}
q_{0}=
\begin{cases}
-0.477 \pm 0.391,\,& z_{t} \sim 0.31~~ \mathrm{for~~} \Delta_{f} = 0.4, \\
-0.652 \pm 0.185, \,& z_{t} \sim 0.38~~  \mathrm{for~~} \Delta_{f} = 0.6, \\
-0.645 \pm 0.091, \,& z_{t} \sim 0.51~~ \mathrm{for~~} \Delta_{f} = 1.0 \,.
\end{cases}~~~
\end{eqnarray}
Just for comparison, the WMAP $\Lambda$CDM model plotted in Fig.~\ref{fig:Qwmap}
has $q_{0} = -0.601$ and $z_{t} = 0.77$, while the WMAP quiessence model $q_{0} = -0.704$ and $z_{t} = 0.74$. In this sense,
the reconstruction at low redshifts seems to be consistent with such cosmological models, so we have another proof of
the consistency of the method in that regime.

\begin{figure}[ht]
\centering
\includegraphics[width=8.5cm]{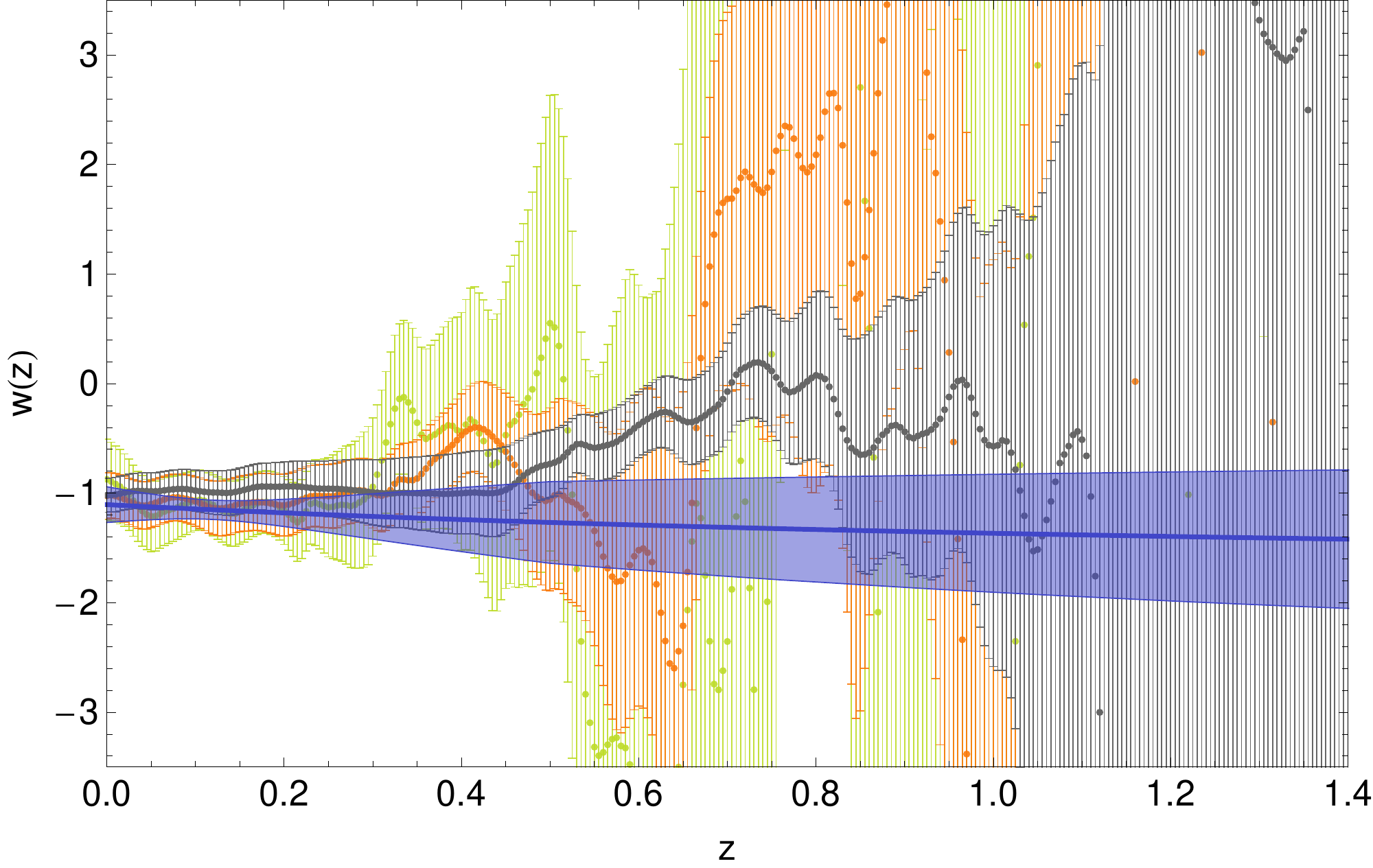}
\caption{Second order polynomial reconstruction with SNeIa (Union2): reconstruction of the dark energy EoS $w(z)$. Different colors are used for different fitting window $\Delta_{f}$: from lightest to darkest for $\Delta_{f}=0.4$, $\Delta_{f}=0.6$ and $\Delta_{f}=1.0$ respectively. Blue contours correspond to the best fits for CPL from real data in
\cite{Sendra:2011pt}.}\label{fig:EoSmap}
\end{figure}

\paragraph{EoS parameter}
It is not an easy task to argue whether  the EoS parameter $w(z)$ has a constant or a dynamical behaviour (as it is possible to verify by
having a look at Fig.~\ref{fig:EoSmap}):
for high redshifts the noise and the errors make it impossible to draw any strong conclusion.
Nevertheless, we can offer estimates for the present values:
\begin{eqnarray}
w_{0} =
\begin{cases}-0.874 \pm 0.369 & \mathrm{for~} \quad \Delta_{f} = 0.4, \nonumber \\
-1.025 \pm 0.211 &\mathrm{for~} \quad \Delta_{f} = 0.6,  \\
-1.019 \pm 0.154 & \mathrm{for~} \quad \Delta_{f} = 1.0\,.
\end{cases}~~~
\end{eqnarray}
Note that those are all compatible with both the WMAP $\Lambda$CDM and quiessence model.
Another interesting value is $w_{0.5}=w(z=0.5)$: following \cite{Wang:2008zh} it is worth considering
an alternative parametrization for the dark energy EoS, as opposed to the most used \cite{Albrecht06}
Chevalier-Polarski-Linder (CPL) model\footnote{As it is well known, the CPL parametrization corresponds to $w(z)=w_0+w_a(1-a)$.} \cite{Chevallier01,Linder03}. This scenario (specifically, a reparametrization
of the former) addresses the values $w_{0}$ and $w_{0.5}$,
instead of the asymptotic value $w_{a} = w (z \rightarrow \infty)$ (see footnote). This pair of values has the advantage of lower correlation
and informs us about a redshift range well accessible from current data. The reconstructed estimations for this alternative
second dark energy parameter are:
\begin{eqnarray}
w_{0.5} =\begin{cases} ~~0.551 \pm 2.079 & \mathrm{for~~} \quad \Delta_{f} = 0.4, \\
 -1.058 \pm 0.835 & \mathrm{for~~} \quad \Delta_{f} = 0.6,  \\
-0.733 \pm 0.307 & \mathrm{for~~} \quad \Delta_{f} = 1.0 \;. \end{cases}
\end{eqnarray}

This result shows
more efficiently than the previous ones
how  difficult it is to extract information from
small fitting windows. Moreover, the largest fitting window width we have chosen is again quite free of biasing
effects, and it gives values that are in agreement with those for the smallest window. The values corresponding to
$\Delta_{f}=0.6$ seem to be
consistent with a constant EoS, even though the errors on $w_{0.5}$ are really large. On the other hand,
results from $\Delta_{f}=1.0$ are more compliant with  a varying EoS. Both are also consistent with the results in \cite{Escamilla11},
where the SNeIa Union2 sample was fitted with the $(w_{0},w_{0.5})$ model. This is also a possible confirmation
that the reconstruction works fine, at least up to $z \approx 0.4-0.5$.

\subsection{GRBs}

As discussed in the previous related section, the GRBs sample to be used is similar to the one
studied in \cite{DalyDjorgovski3}, the difference being it is calibrated so that the GRBs Hubble diagram
coincides with the SNeIa one in the redshift range where they overlap.

We have performed a cut similar to the one we did for the SNeIa sample: the original
data set was approximately homogeneous until a $z\approx5$ and shows only $2$ GRBs for higher
redshift up to $z \approx6.6$. Trying to fit these data is pointless, and all one gets are  large noisy
trends at their related redshift; for this reason we perform fits without considering these last two points.

The minimum length that can be used with GRBs for which the number of datapoints is higher
than the number of parameters is $\Delta_{f}= 1.3$. Nevertheless,  this is a border line case, because the number of
datapoints is far too low and in many cases exactly equal to the number of fitting parameters, so that
eventual results are statistically invalid. This also gets clearly reflected in the error
bars, which are rather big and give no information at all about the trends of the quantities.
The minimal number requirement we have also mentioned before (having $\geq 10$ points in each fitting window)
is first fulfilled by $\Delta_{f} = 3.0$. For further insight, we also present data from
a wider fitting window with $\Delta_{f} = 4.0$, while one must always keep in mind all the problems
that enlarging $\Delta_{f}$ way too much brings about.

\begin{figure*}
\begin{tabular}{cc}
\includegraphics[width=7.cm]{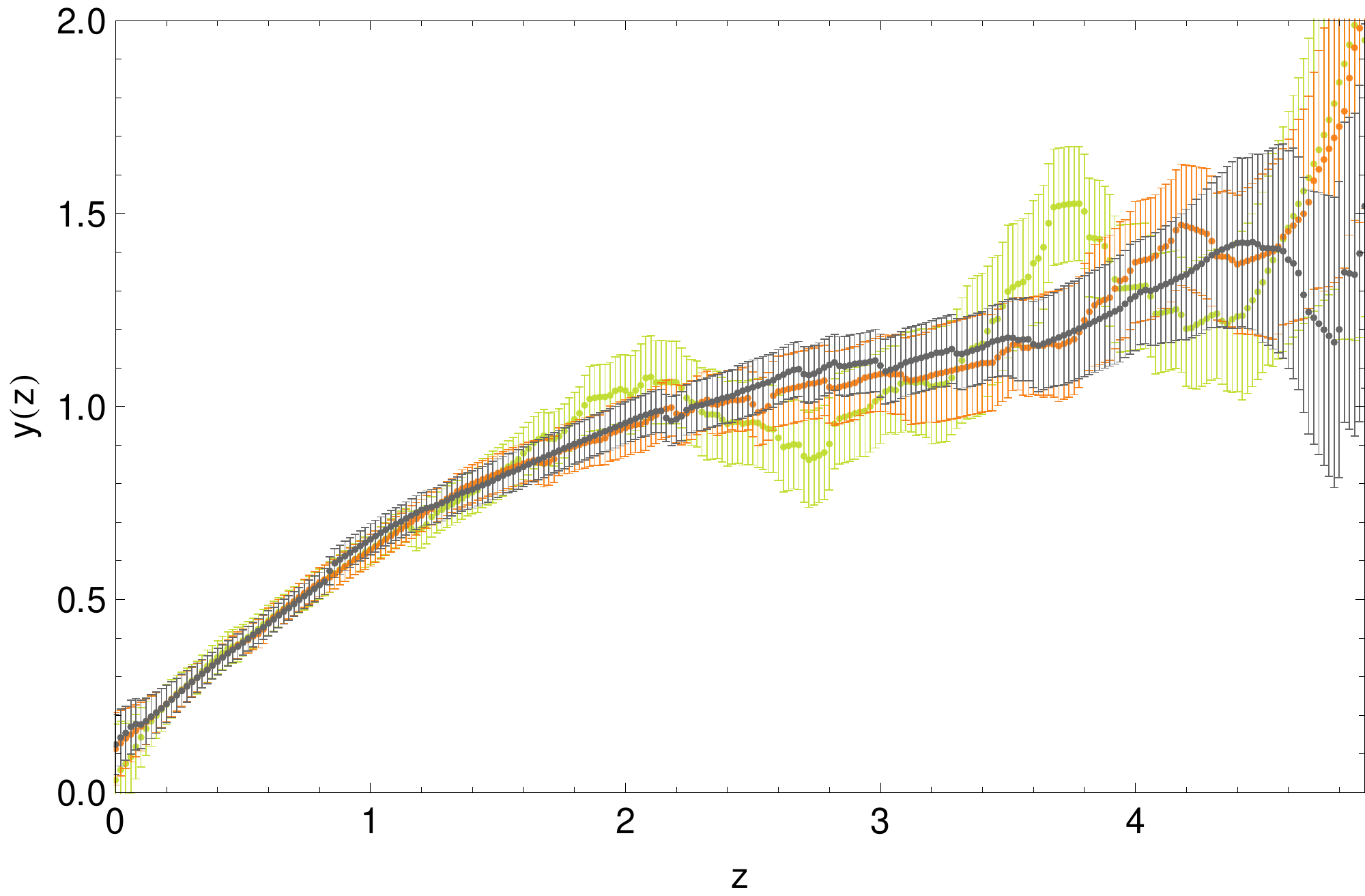}&
\includegraphics[width=7.cm]{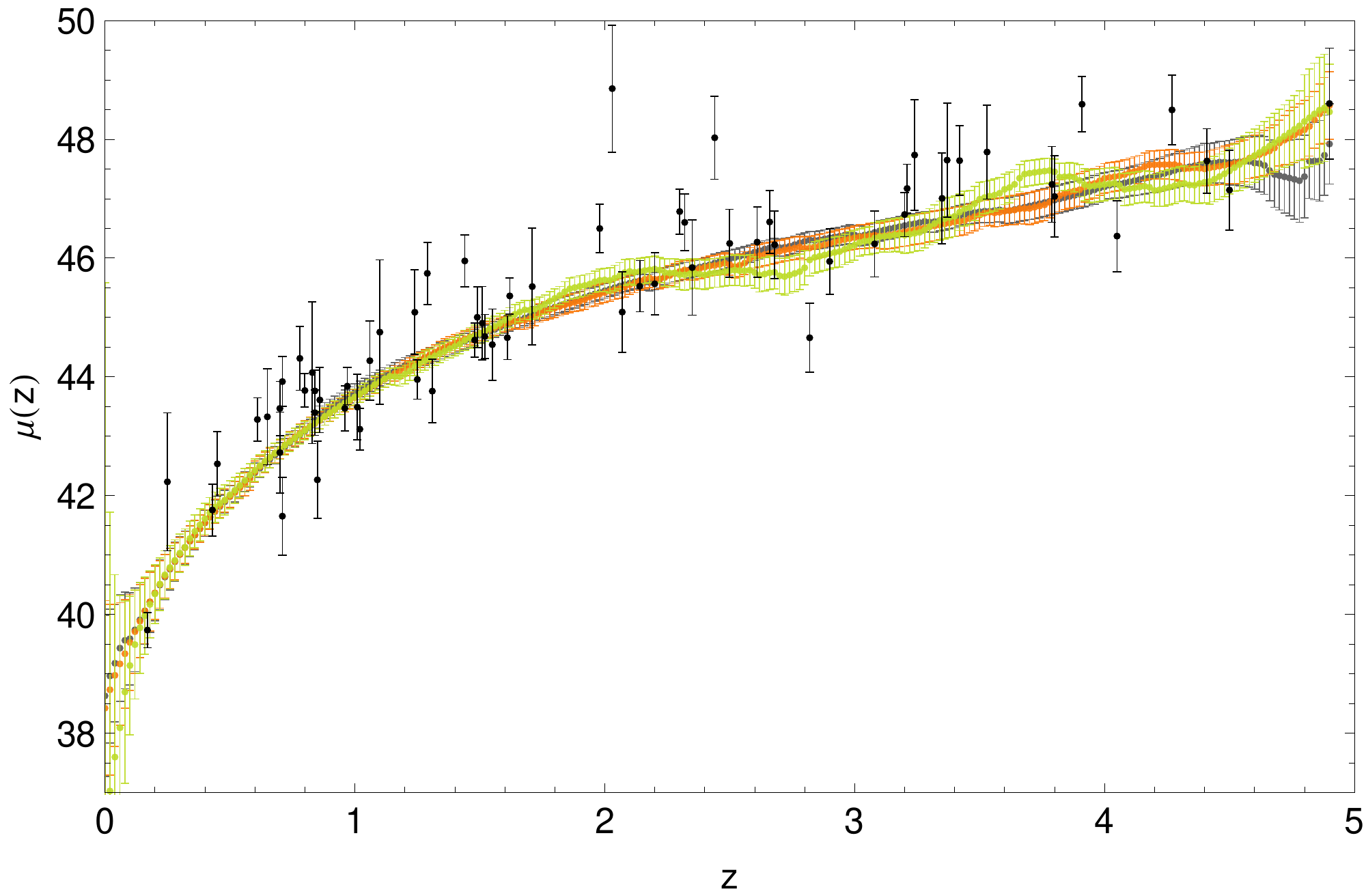}\\
(a)&(b)\\
\includegraphics[width=7.cm]{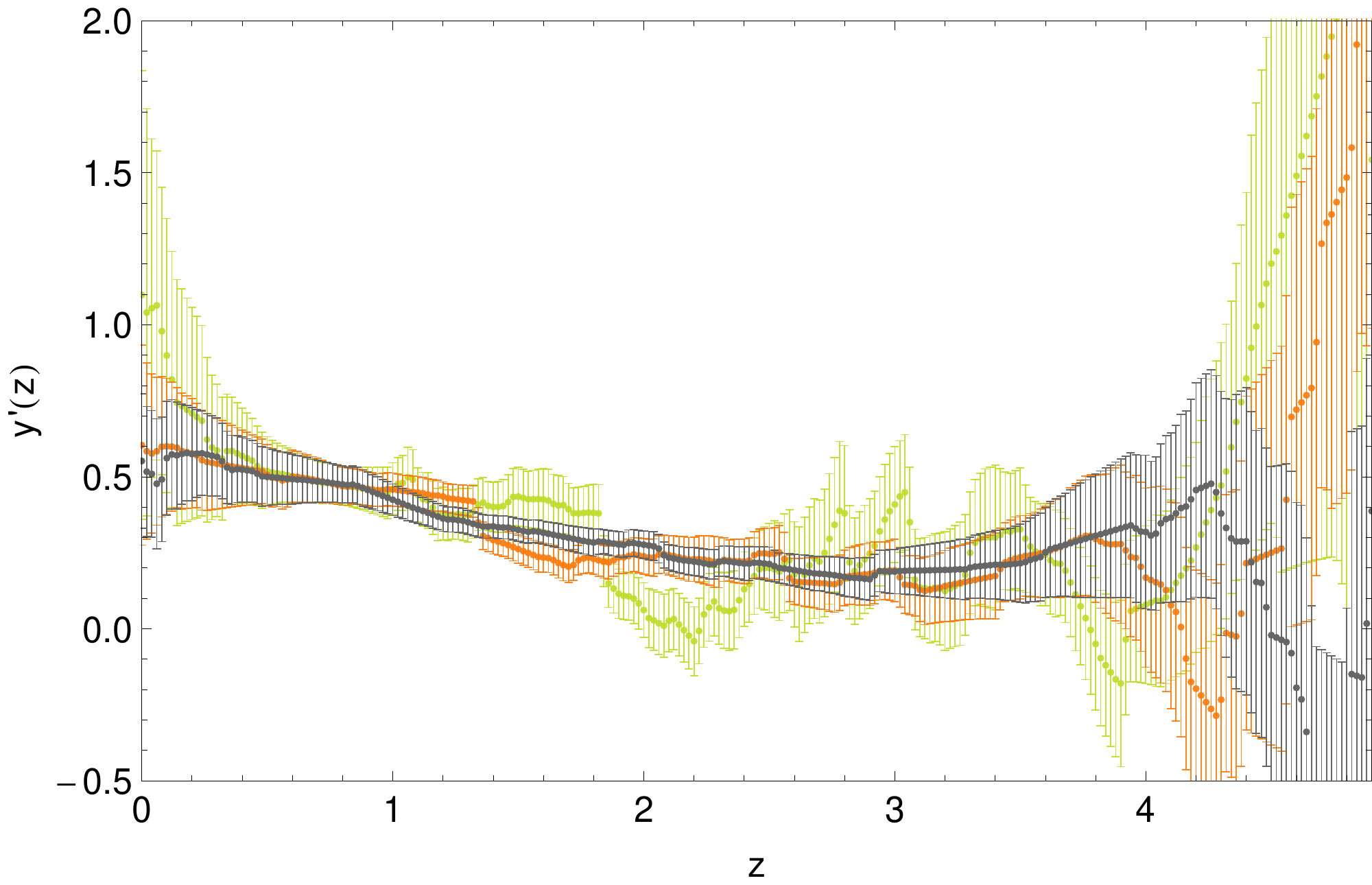}&
\includegraphics[width=7.cm]{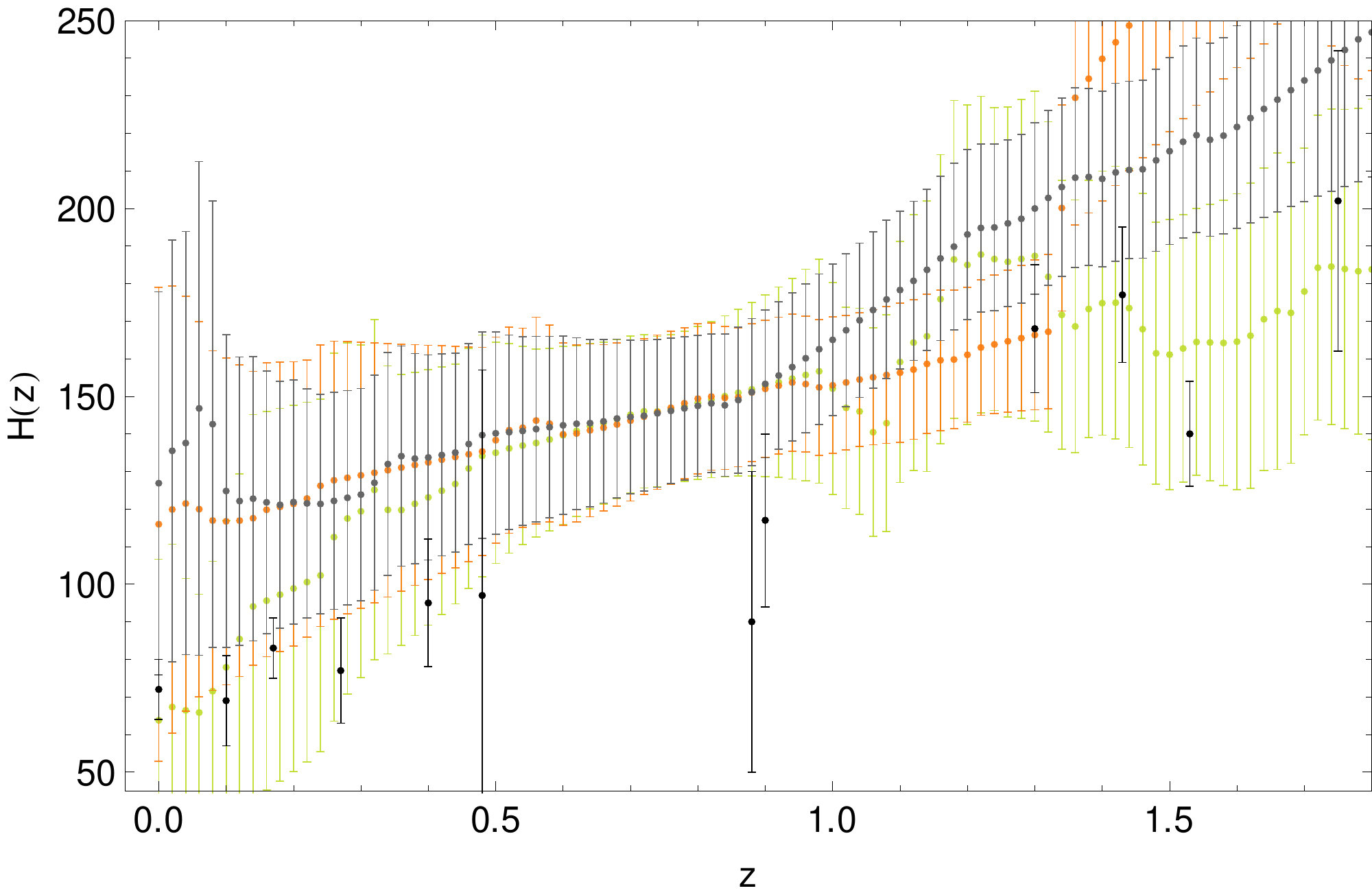}\\
(c)&(d)\\
\includegraphics[width=7.cm]{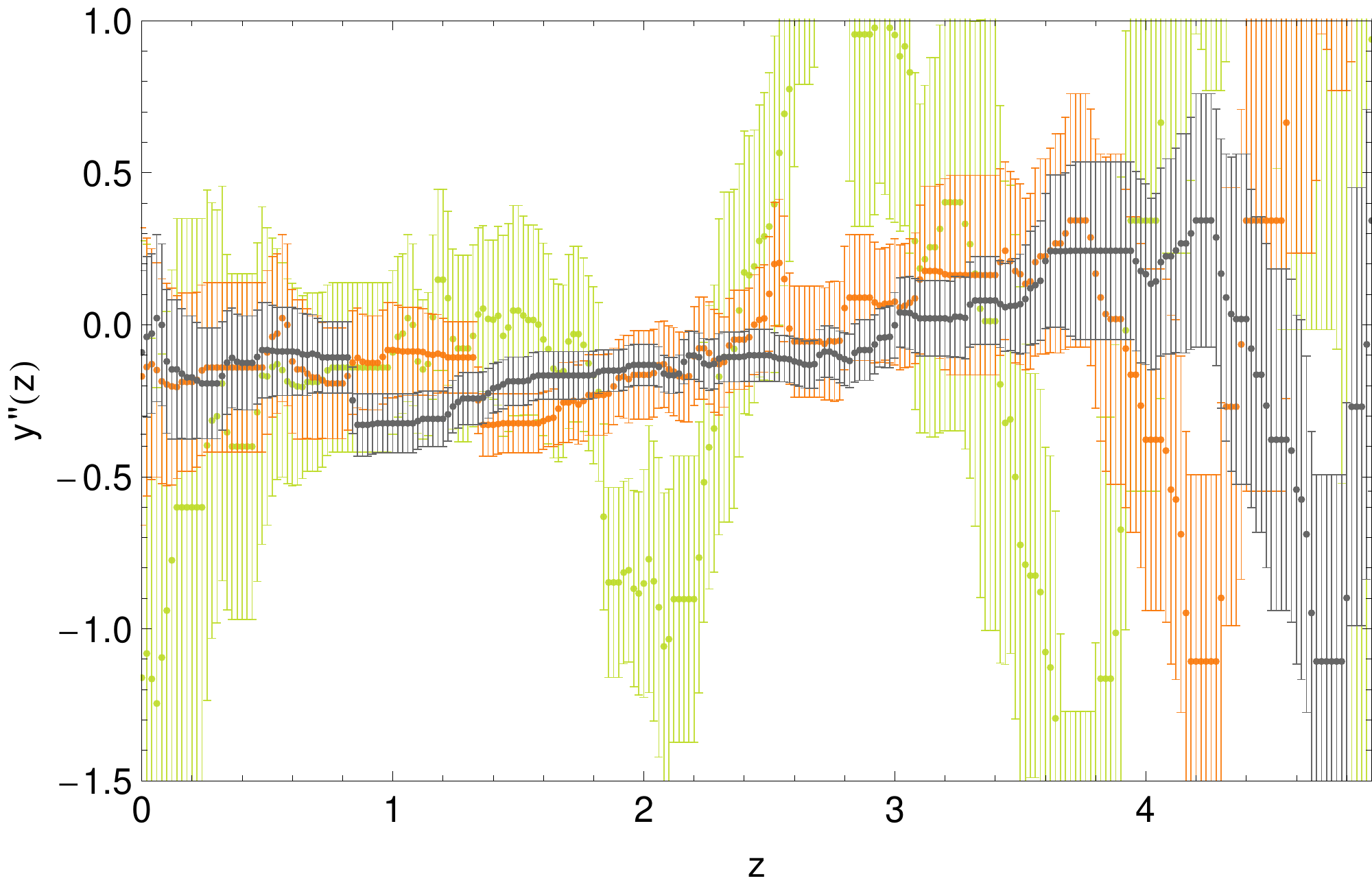}&
\includegraphics[width=7.cm]{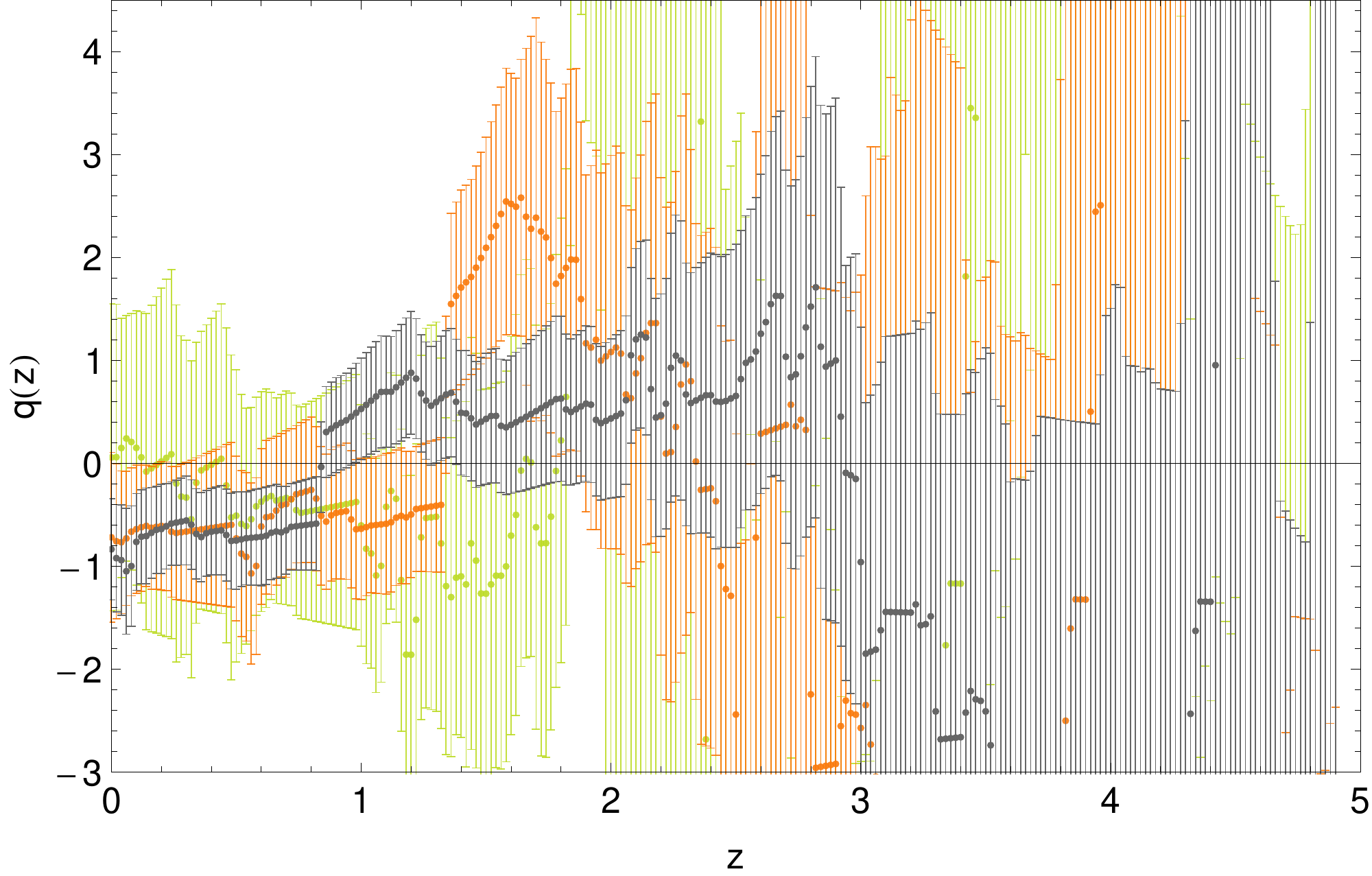}\\
(e)&(f)\\
\end{tabular}
\includegraphics[width=7.cm]{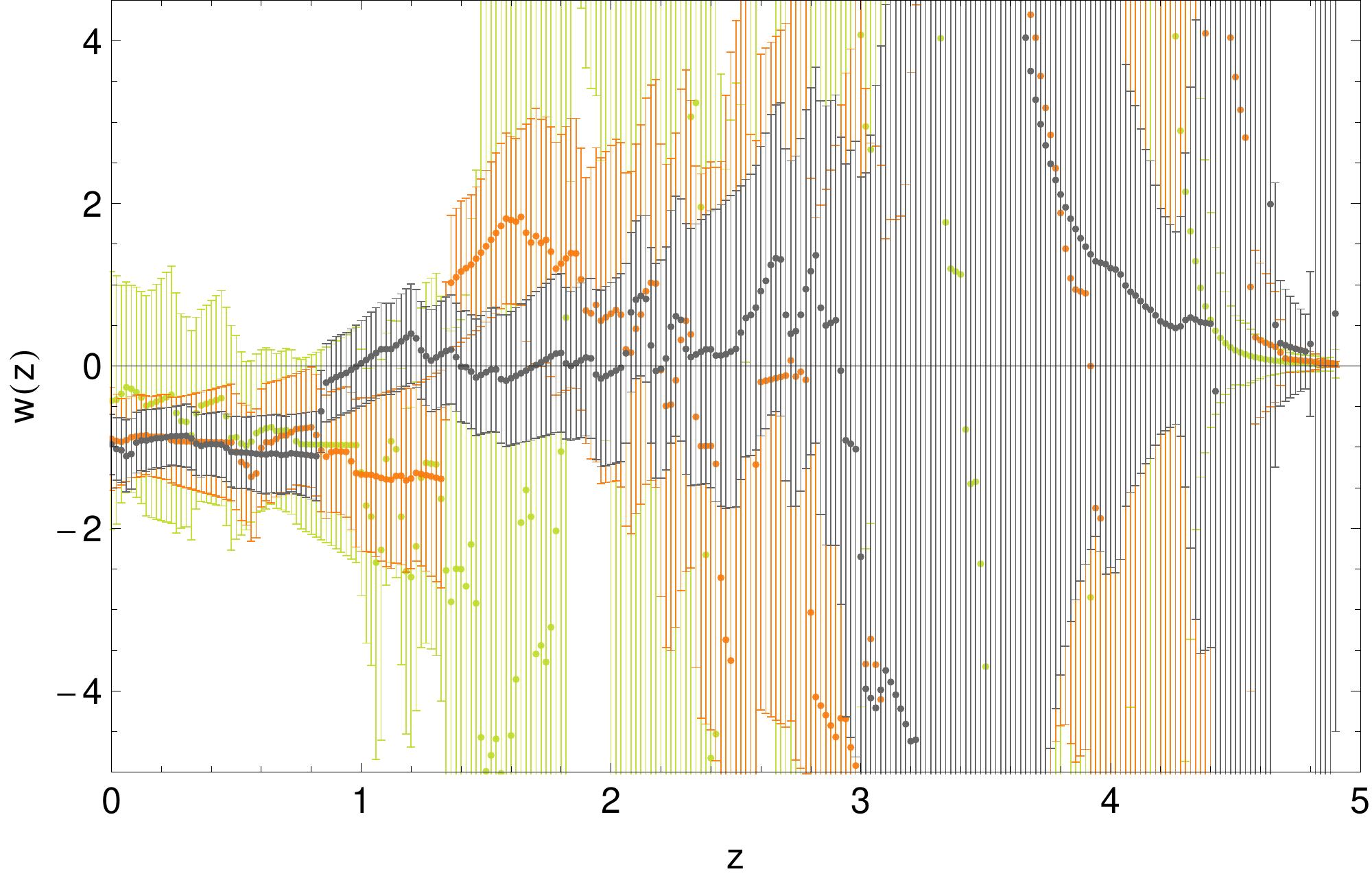}\\
(g)
\caption{Second order polynomial reconstruction with GRBs (Schaefer-Cardone). Coordinate distance $y(z)$ and its derivatives ($y'(z)$ and $y''(z)$), and the derived cosmological quantities: distance modulus $\mu(z)$, Hubble function $H(z)$, deceleration parameter $q(z)$ and the dark energy EoS $w(z)$. Different colors are used for different fitting window $\Delta_{f}$: from lightest to darkest for $\Delta_{f}=3$, $\Delta_{f}=4.$ respectively.}
\label{fig:parameters2PG}
\end{figure*}

In Fig.~\ref{fig:parameters2PG} we can see the reconstructed variables $y(z)$, $y'(z)$, $y''(z)$ for the GRBs sample.
The reconstructed $y(z)$ seems to be conditioned by the low number of data points in the sample, which translates
into oscillations  and growing error bars at high redshifts, larger than those  for the SNeIa reconstruction.
The first order derivative $y'(z)$ presents two main differences with respect to the SNeIa case: $1.$ it decreases faster,
and most importantly, $2.$ its present value is sensibly lower. This will have a strong impact in the
reconstruction of $H(z)$, which has a strong dependence  on $y'(z)$ and becomes manifest in Fig.~\ref{fig:parameters2PG}(d).
This quantity displays a large deviation from current data, and it results in  a very large Hubble constant value (with
a large uncertainty as well):
\begin{eqnarray}
H_{0}&=\begin{cases} 115 \pm 63 & \mathrm{for~~} \quad \Delta_{f} = 3.0,  \\
 126 \pm 51 & \mathrm{for~~} \quad \Delta_{f} = 4.0\,. \end{cases}
\end{eqnarray}
These values clearly differ from current accurate estimates from other astronomical probes \cite{Riess:2011yx},
and no  good agreement is found with \cite{Stern10} as the global trend is concerned.

When considering the second order derivative $y''(z)$, represented in Fig.~\ref{fig:parameters2PG}(e),
one can notice the abrupt change at $z \approx 0.8$, which is also reflected in the related parameters, i.e. $q(z)$ and $w(z)$,
as shown in Figs.~\ref{fig:parameters2PG}(f) and \ref{fig:parameters2PG}(g). The deceleration and the EoS parameter trends are conditioned by very large error bars; the $w$ profile,
in particular, seems to be compatible with a constant EoS, $w \sim -1$ for $z \leq 0.8$ and $w \sim 0.1$ for
$0.8\leq z \leq 2.$; but also becomes divergent at $z \approx 3.5$ due to the denominator entering its expression.

The observed large discrepancy between the GRBs and the SNeIa reconstruction
is probably both
 due to the low number of GRBs data in this regime and to intrinsic
 shortcoming in the calibration laws required for the treatment of GRBs datapoints.
These disadvantage make us feel  the GRBs are not reliable to be used in this model-independent framework. However, we expect that
an improvement on the knowledge of GRBs can probably lead to tighter measures and then to a better implementation in this method.

\subsection{Mock}
\label{sec:Mock}

The main question arising from the previous section is:
what is the degree of confidence on the reconstruction
of cosmological quantities with this method? Put in other words,
is there any way to discriminate and disentangle the ``real'' reconstruction from
noisy features caused by shortcomings in the algorithm? And considering the particular kind of data
we have used, this can also be translated into another question: until what redshift is the reconstruction
well based?

To try to throw some light into these matters, we have used a mock SNeIa sample: this is built on a fiducial cosmological model,
so that, in theory, if the method works well, one should expect to obtain reconstructed quantities compatible with this model.
To create SNeIa mock samples we have taken into account the specifications of the
Wide-Field Infrared Survey Telescope (WFIRST) \cite{wfirst}, a space mission planned by NASA, which has among its
primary objectives to explore the nature of dark energy employing three distinct techniques: measurements of baryon acoustic oscillations,
SNeIa distances, and weak gravitational lensing.
The  redshift distribution of the expected  SNeIa to be observed within various redshift bins
in a survey of that sort is reported in \cite{JDEM}.
The main redshift range extends up to $z\sim 1.6$, with $4$ expected SNeIa at $z>1.6$. The  very
low redshift subsample, i.e. the $500$ SNeIa at $z<0.1$, is supposed to come from the Nearby Supernova Factory
(SNfactory) \cite{SNfactory} with measurements in the redshift range $(0.03,0.08)$, so we are going
to assume the former as the lowest redshift value in our mock sample.

Our formulae for errors on SNeIa magnitudes stems from a recipe used in the binning approach \cite{snerror},
which we have however adapted for its application to non-binned data. According to this prescription errors are
calculated as follows :
\begin{equation}\label{eq:error}
\sigma_{m}^{\mathit{eff}} = \sqrt{\sigma_{int}^2+ \sigma_{pec}^2 + \sigma_{syst}^2} \; ,
\end{equation}
where: $\sigma_{int} = 0.15$ is the intrinsic dispersion in magnitude per SNeIa, assumed to be constant and independent of redshift for all  well-designed surveys; $\sigma_{pec} = {5 \sigma_{v}}/{\ln(10) c z}$ is the error due to the uncertainty in the SNeIa peculiar velocity, with $\sigma_{v} = 500$ km$/$s, $c$ the light velocity and $z$ the redshift for any SNeIa; $\sigma_{syst} = 0.02 (1+z)/(1+z_{max})$ is the floor uncertainty related to all the irreducible systematic errors with cannot be reduced statistically by increasing the number of observations. The value $0.02$ comes from the fact that WFIRST is a space-based mission (it would be $0.05$ for terrestrial-based observations); $z_{max}$ is the maximum observable redshift in the considered mission and this linear term in redshift is used to account for the dependence with redshift of many of the possible systematic error sources (for example the Malmquist bias or gravitational lensing effects).
The fiducial model we have used is  the  wcdm+sz+lens case from WMAP7-year
\cite{wcdm}, which has $h = 0.75^{+0.15}_{-0.14}$,
$\Omega_m= 0.259^{+0.099}_{-0.095}$,
and is phantom-like with
$w= -1.12^{+0.42}_{-0.43}$.

We will use a mock sample generated with that prescription to study the relation of the proposed reconstruction method with
three different features: $1.$ data dispersion;  $2.$ changes in the fitting window length and $3.$ the range of
validity of the method.

The first point is addressed starting from the algorithm used to generate the
mock sample. We take a set of values of the cosmological parameters in the fiducial model which comprises
these cases: the best fit, the best fit plus the $1\sigma$ error, and the best fit minus the $1\sigma$ error.
We then construct all possible combinations of them, and then for these sets, and for any
redshift in the sample, we find the largest possible variation in the distance modulus. We store this set of
values as they are going to serve as our dispersion seeds. The next step is to build four mock samples as follows.
We take the distance modulus corresponding to each redshift according to the fiducial model, and then we add
to it a random value extracted from a Gaussian centered at zero and with a dispersion which is a number of times
($0,0.25,0.5$ or $0.75$)
the corresponding value from our dispersion seeds vector. The first case
corresponds to  mock data points that perfectly resemble the exact fiducial model,
while he last value gives
a more realistic dispersion similar to the from the present Union2 sample.

For the sake of clarity, we will show results for a window length $\Delta_{f} = 0.6$, which is the best case
for the Union2 sample. In fact, it should also fare well with WFIRST (recall that with a sample with more points and more dense, the fitting
window could be diminished with equally good final results). Moreover, we will restrict the discussion to the deceleration and EoS parameter,
which are the two most sensitive parameters to eventual features.

\begin{figure*}
\begin{tabular}{cc}
\includegraphics[width=7.cm]{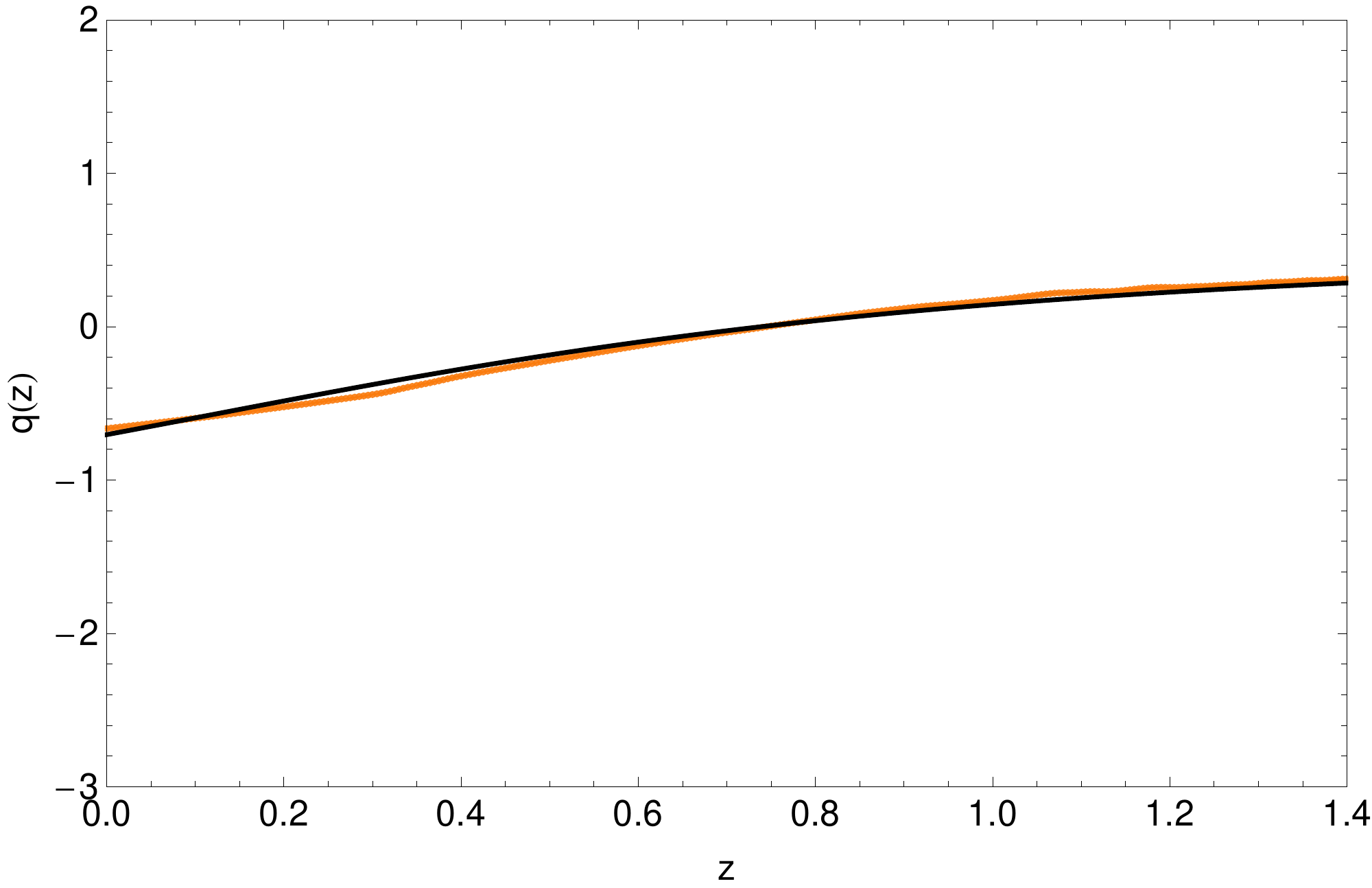}&
\includegraphics[width=7.cm]{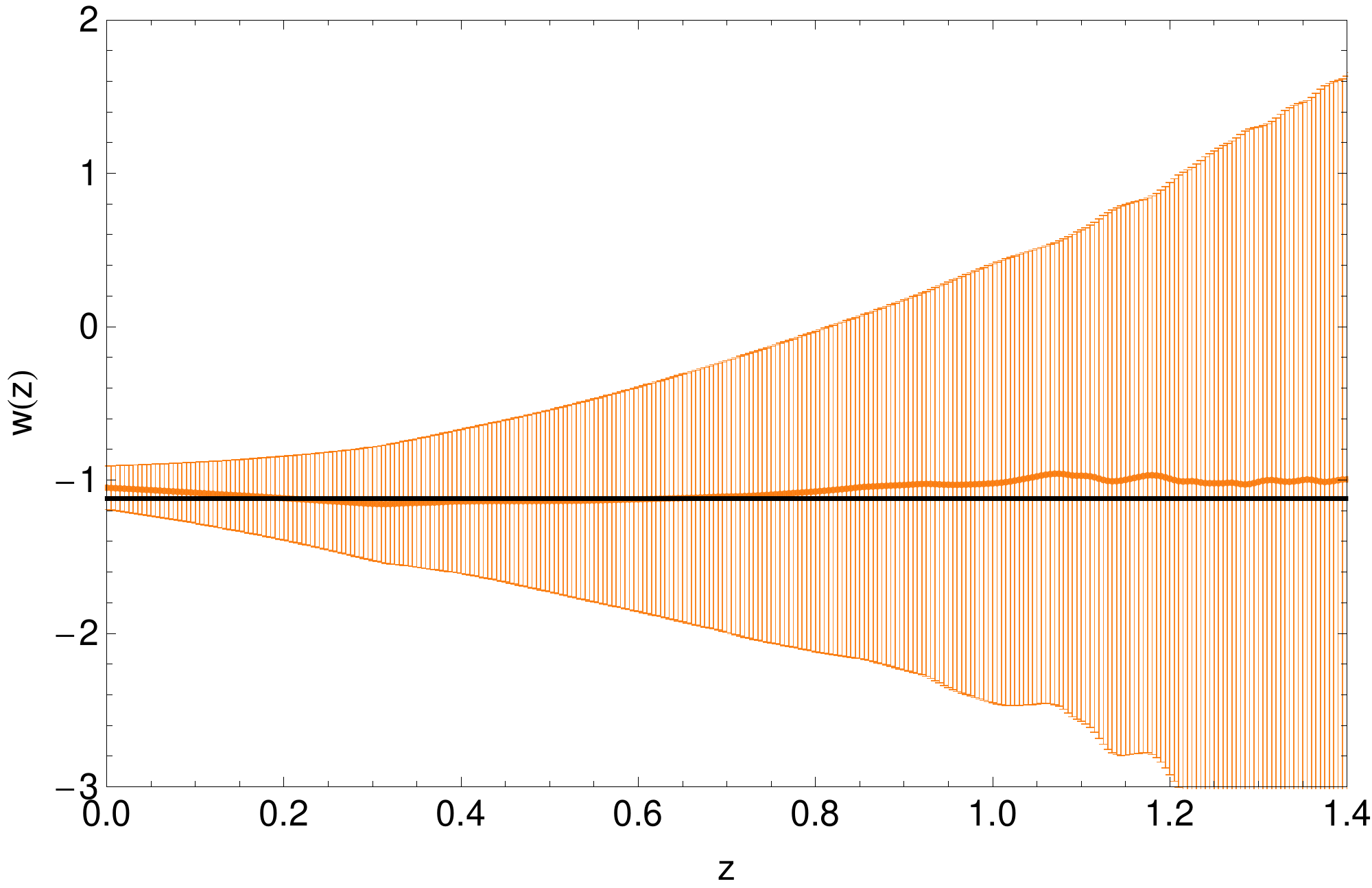}\\
(a)&(b)\\
\includegraphics[width=7.cm]{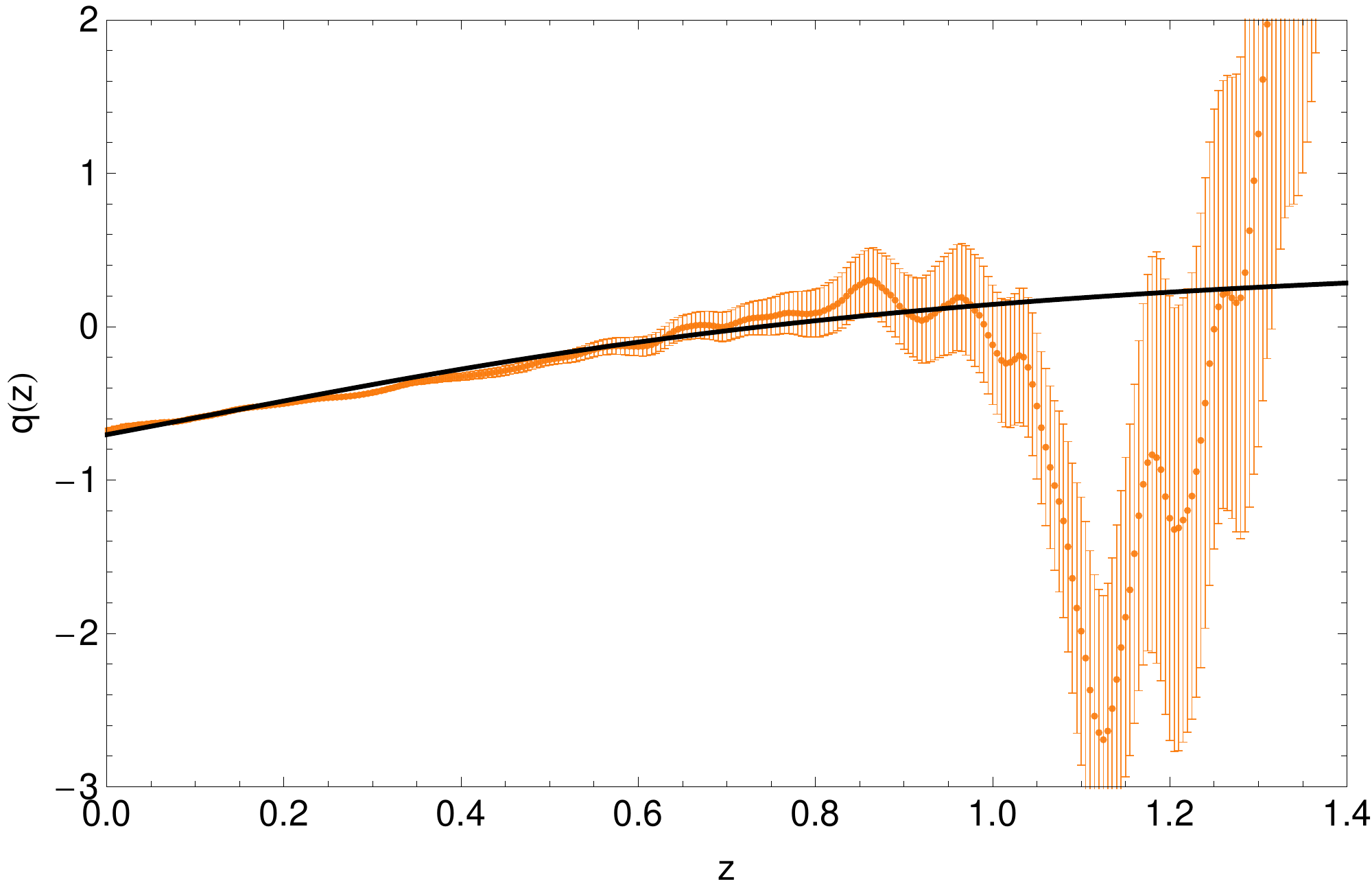}&
\includegraphics[width=7cm]{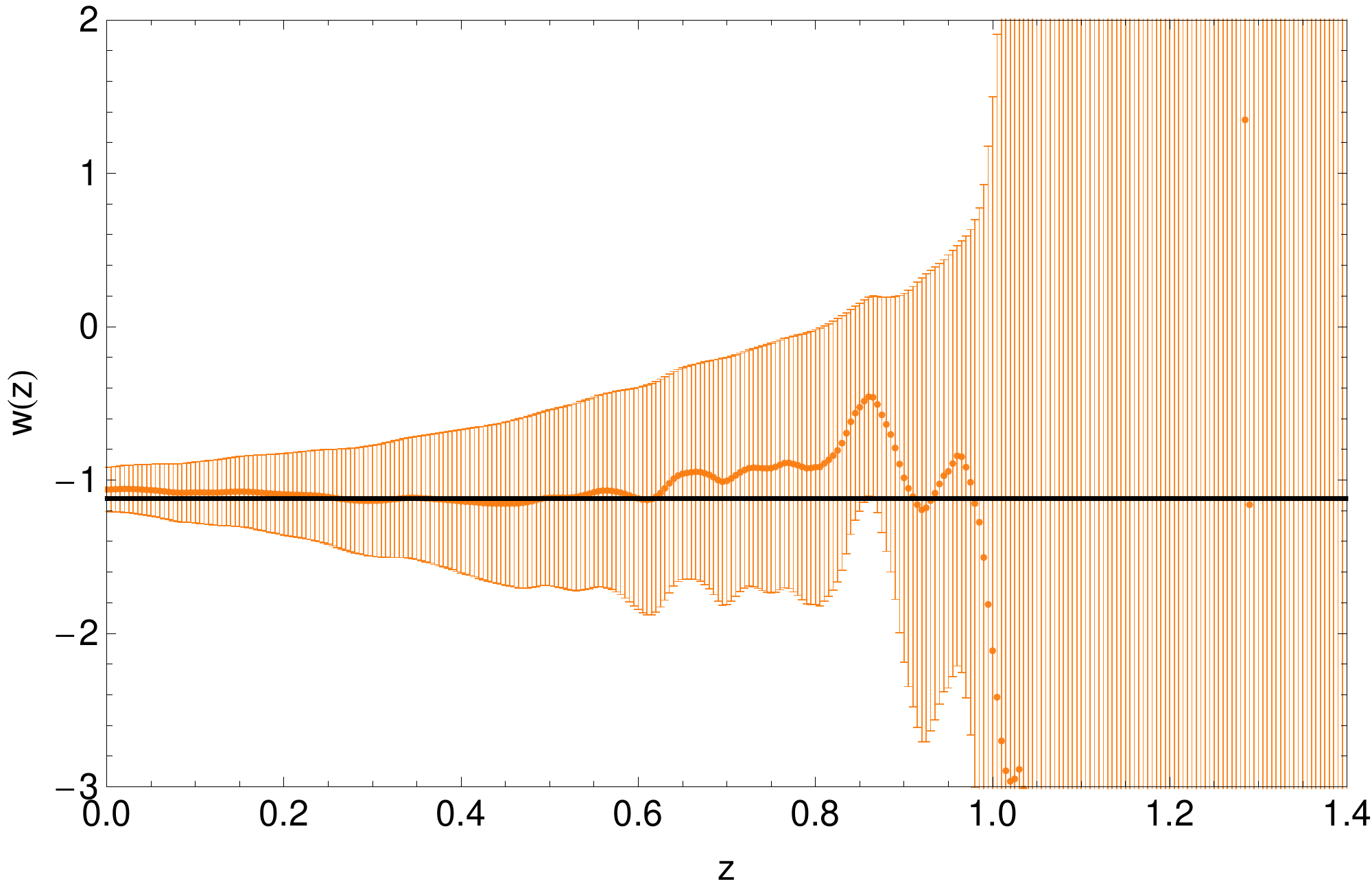}\\
(c)&(d)\\
\includegraphics[width=7.cm]{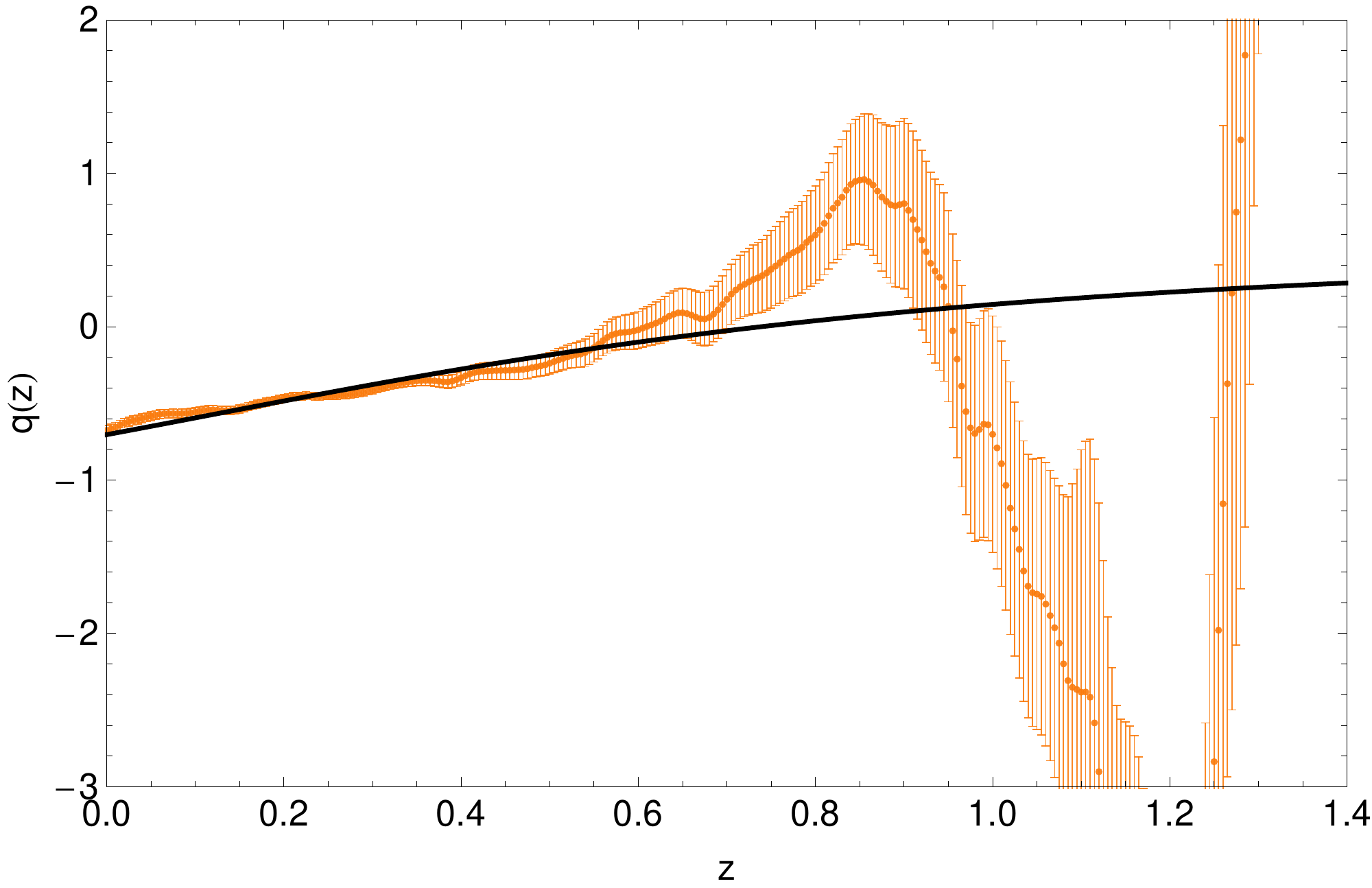}&
\includegraphics[width=7.cm]{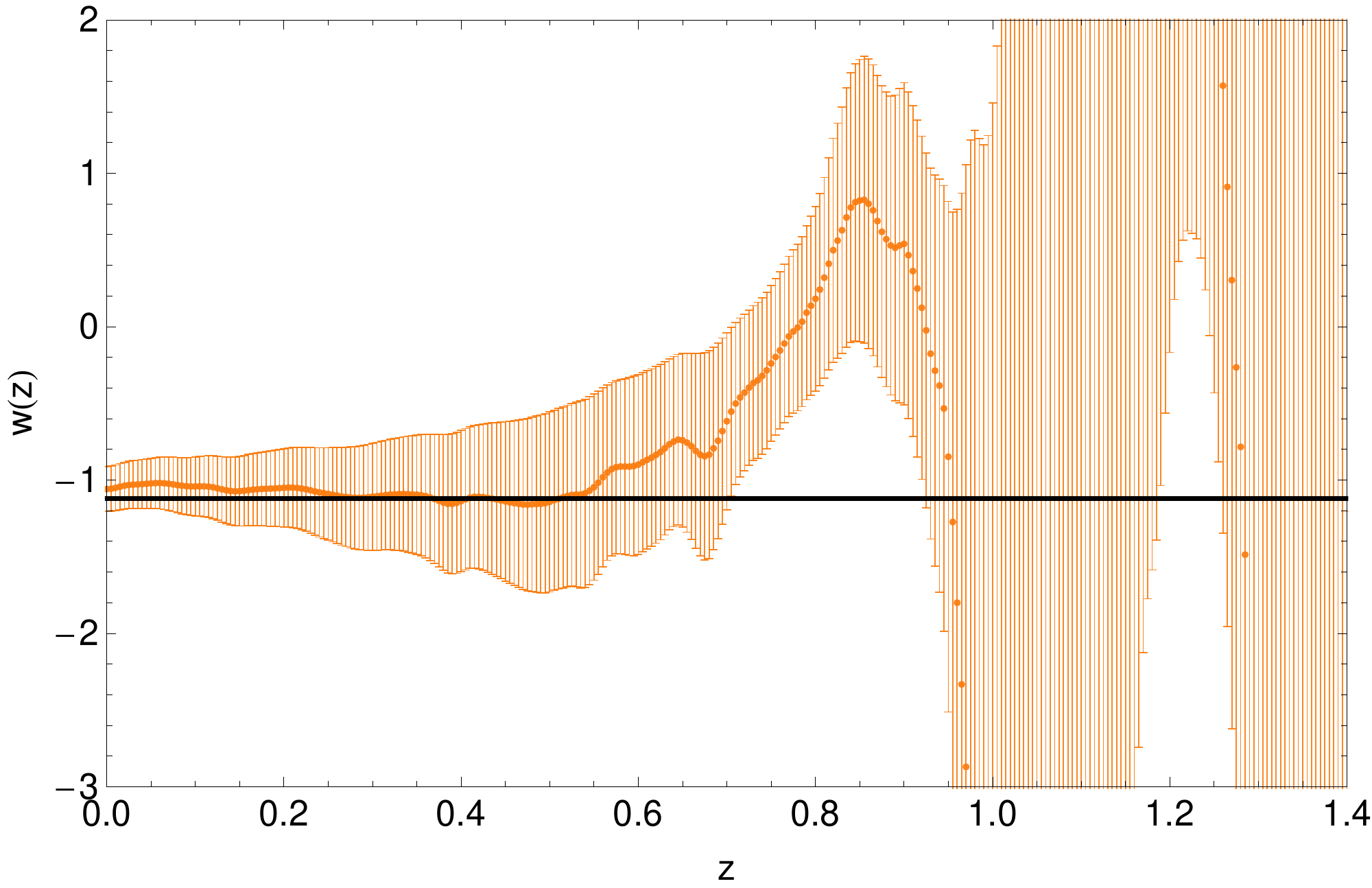}\\
(e)&(f)\\
\includegraphics[width=7.cm]{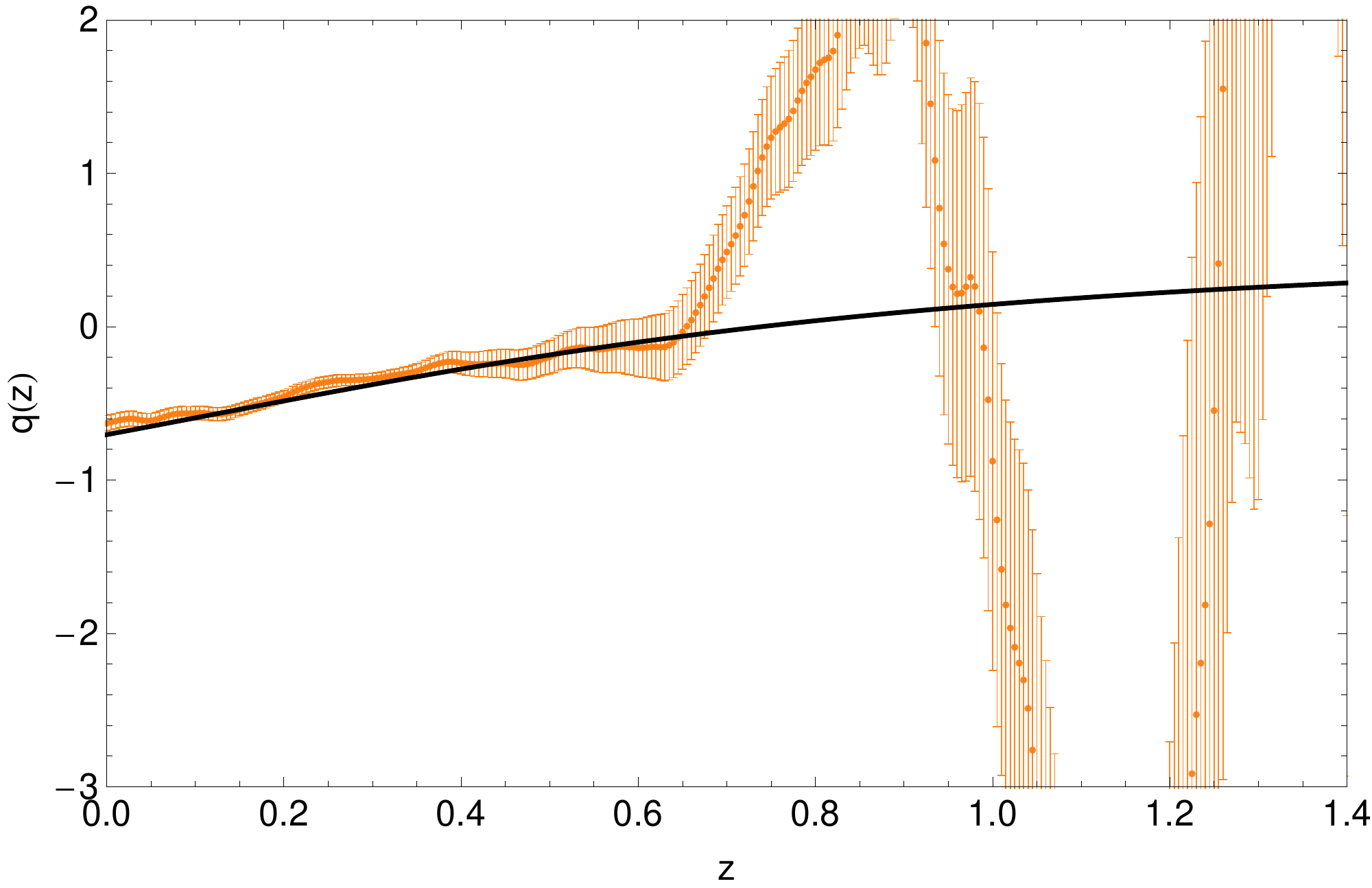}&
\includegraphics[width=7.cm]{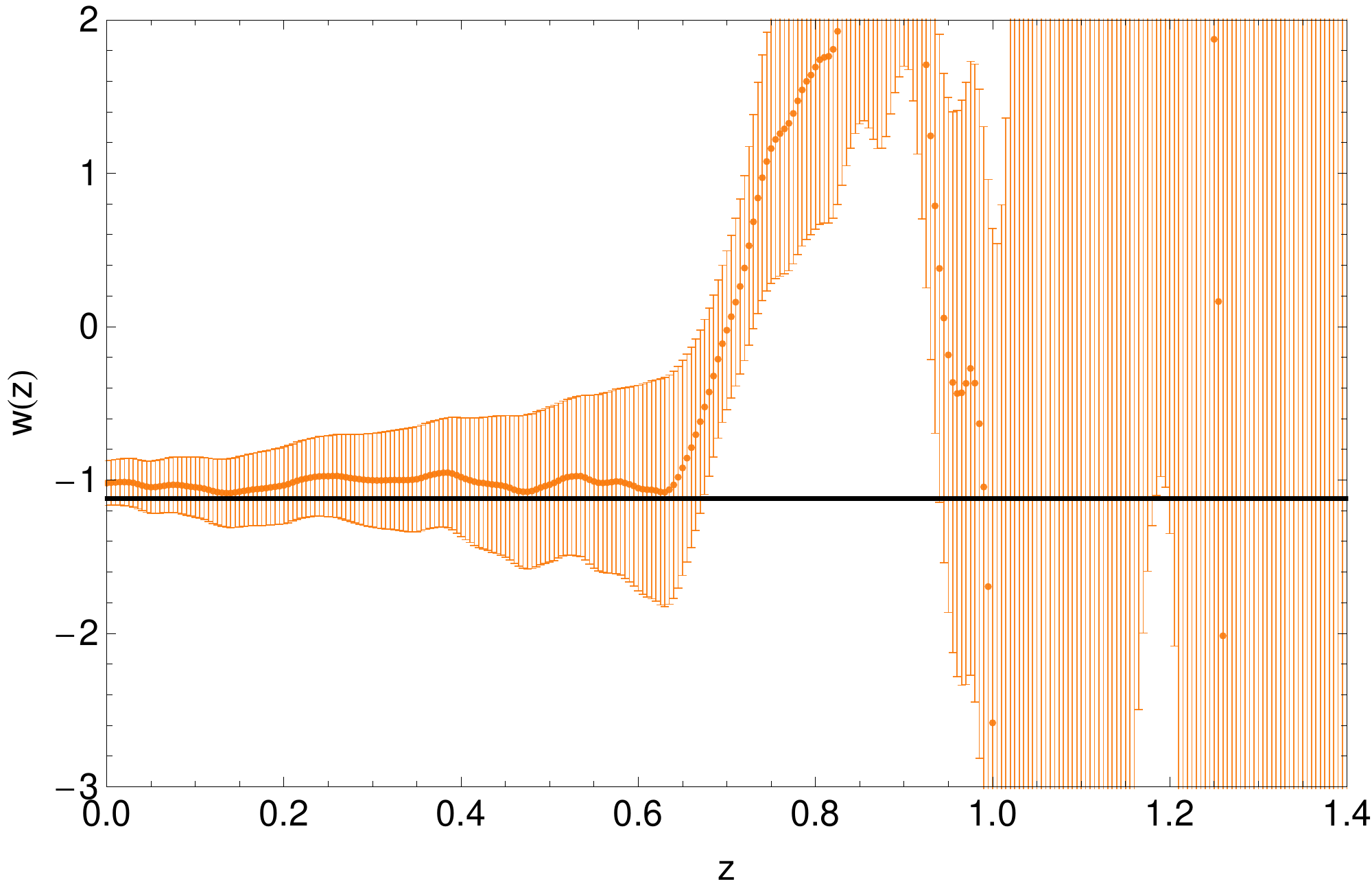}\\
(g)&(h)\\
\end{tabular}
\caption{Deceleration $q(z)$ and dark energy EoS $w(z)$ - second order polynomial with WFIRST mock data with $\Delta_{f}=0.6$ and $\sigma_g=0.02$. Black lines are the considered quantities in the fiducial cosmological model used to create the mock data set, the  wcdm+sz+lens case from WMAP7-year
\cite{wcdm}, with $h = 0.75^{+0.15}_{-0.14}$, $\Omega_m= 0.259^{+0.099}_{-0.095}$ and is phantom-like with $w= -1.12^{+0.42}_{-0.43}$. \textit{First panel:} Null dispersion mock data. \textit{Second panel:} Mock  data dispersion factor $0.25$. \textit{Third panel:} Mock  data dispersion factor $0.5$.
\textit{Fourth panel:} Mock  data dispersion factor $0.75$. Dispersion values are as described in the related section.}\label{fig:jdem_dispersion}
\end{figure*}

\begin{figure}
\centering
\includegraphics[width=7.cm]{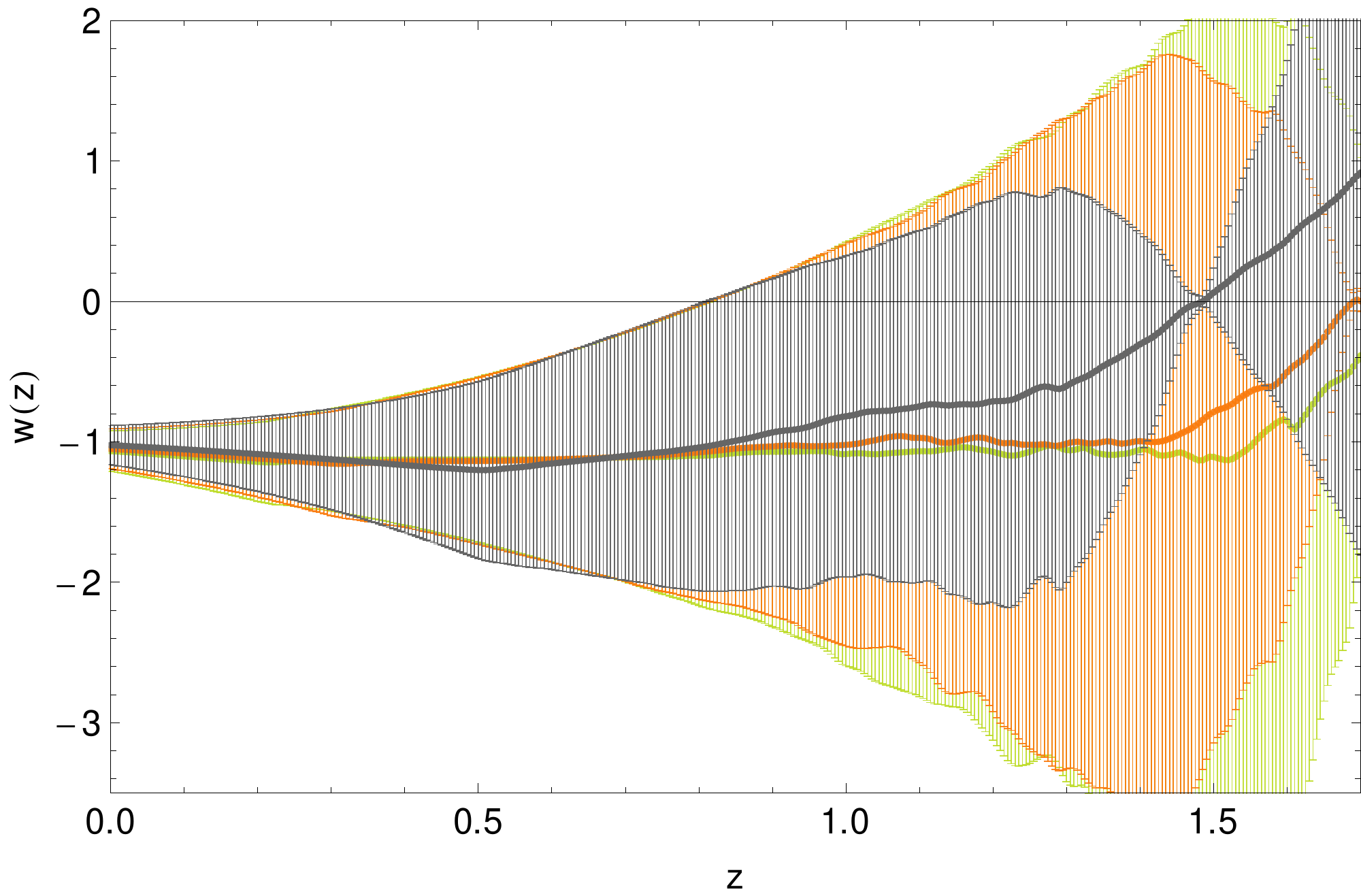}\\
(a)\\
\includegraphics[width=7.cm]{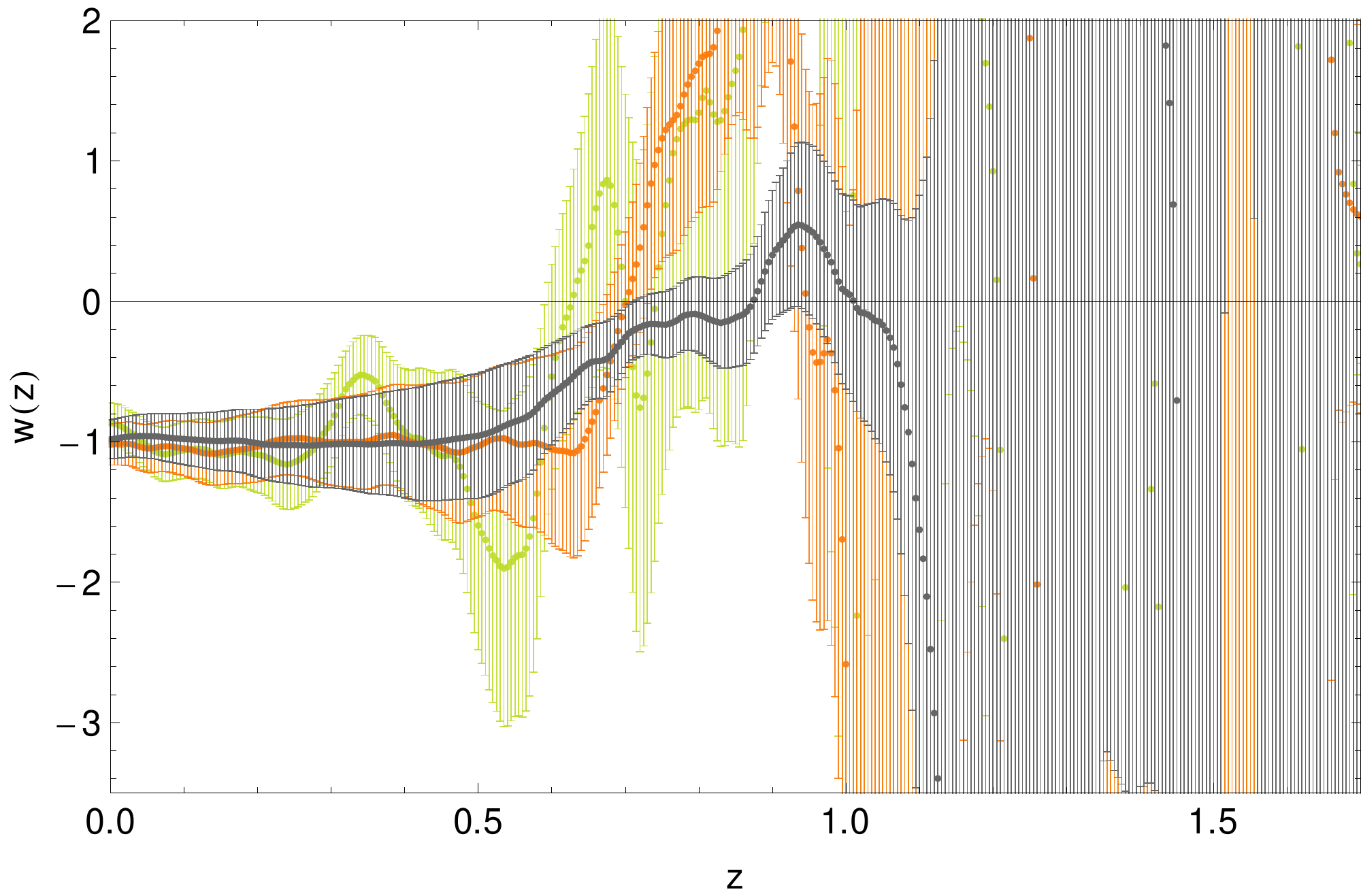}\\
(b)
\caption{Dark energy EoS function $w(z)$ - second order polynomial with WFIRST mock data. \textit{Left:} mock data with null dispersion and different windows functions (from lightest to darkest, $\Delta_{f} = 0.4$, $\Delta_{f} = 0.6$ and $\Delta_{f} = 1.0$). \textit{Right:} mock data with dispersion $0.75$ and different windows functions (from lightest to darkest, $\Delta_{f} = 0.4$, $\Delta_{f} = 0.6$ and $\Delta_{f} = 1.0$).}
\label{fig:jdem_window}
\end{figure}

From Figs.~\ref{fig:jdem_dispersion} it is clear that the reconstruction method
depends on the dispersion of the observed data. In fact, an effect of this sort should not come as a surprise,
as there is always an intrinsic floor for this dispersion (depending on the physics of the observed phenomena and on the reduction procedure of observational
data). Consider, for instance the ideal case of null dispersion, as depicted in Figs.~\ref{fig:jdem_dispersion}(a) and \ref{fig:jdem_dispersion}(b), which means that the only element that makes the data differentiating from the fiducial model is the error on the distance modulus. For this case the reconstruction fails at low and
high redshift. As discussed in previous section, at low redshift we have biased estimations of the parameters due to
the asymmetry of the fitting window, while at high redshift we have few points in each fitting window. As long as the
dispersion grows, the deviation from the fiducial model grows too along with the noisy features,
finally reaching the more realistic case ($0.75$). This one is  shown in Figs.~\ref{fig:jdem_dispersion}(g) and \ref{fig:jdem_dispersion}(h), where results are quite similar to those
derived from Union2.  It follows that an improvement on the observational measures will be translated into a better reconstruction of the
parameters under study. However, it has to be taken into account the limitation at high redshift where the density of points is lower.
In Fig.~\ref{fig:jdem_window} we can also verify how the fitting window influences results: in the ideal case of no dispersion
in the data (Fig.~\ref{fig:jdem_window}(a)), effectively the smaller the fitting window, the better the reconstruction, with only small deviation
at the high redshift tail, where the density of points drops. In the more realistic case in Fig.~\ref{fig:jdem_window}(b), where a larger dispersion in the data is present, we can see how the choice of the fitting window length is crucial. A large window gives a small bias in the estimation of the parameters; on the other side, smaller windows give a better result, but the noise makes it difficult to understand what is the best choice and up to what redshift we have the best reconstruction.

\section{Binned Approach}
\label{sec:Binning}

For completeness, we have also carried out an independent analysis using a binned approach
so as to  get independent values of the magnitudes of interest at certain values of the redshifts. This is, of course,
opposed to the work spirit of the previous sections, which were devoted to the study of general trends
of the $H(z)$ or the dark energy properties $q(z)$ or $w(z)$. The procedure we present and discuss
in this section can be used as an independent check for  theoretical models. We must stress, however, that for
this part of our analysis we have considered SNeIa only, given that according to our findings on GRBs they
do not stand at the moment on the same quality grounds, at least in what concerns the reconstruction methods considered
in this paper.

Our starting point has been to divide the SNeIa data sample into different numbers of redshift bins.  Then, using the data points within each of them, we have  obtained
$y(z)$, $y'(z)$ and $y''(z)$. For this task we have followed basically the steps described in the previous section, but with the slight modification of
giving a low probability to the points near the boundaries of the
bin.

\begin{table*}[htbp]
\centering
\begin{tabular}{c|c|c|c|c|c}
\hline \hline
& N & $z_{med}$ & $H(z_{med})$ & $w(z_{med})$ & $q(z_{med})$ \\ \hline
 \multirow{9}{5mm}{\begin{sideways}\parbox{25mm}{7 bins}\end{sideways}}
        & &   &                    &                    \\
&$77$&$0.020$&$72.376\pm4.717$&$-21.507\pm59.869$&$-23.477\pm66.523$\\
&$76$&$0.037$&$89.058\pm6.928$&$-17.654\pm13.774$&$-21.253\pm16.817$\\
&$77$&$0.120$&$74.221\pm2.238$&$-0.528\pm2.985$&$-0.036\pm3.027$\\
&$77$&$0.262$&$84.428\pm6.905$&$-10.709\pm13.375$&$-9.818\pm12.307$\\
&$76$&$0.383$&$72.376\pm4.717$&$-21.507\pm59.869$&$-23.477\pm66.523$\\
&$77$&$0.554$&$89.058\pm6.928$&$-17.654\pm13.774$&$-21.253\pm16.817$\\
&$76$&$0.872$&$74.221\pm2.238$&$-0.528\pm2.985$&$-0.036\pm3.027$\\
        & &   &                    &                    \\ \hline
 \multirow{8}{5mm}{\begin{sideways}\parbox{25mm}{ 6 bins}\end{sideways}}
        & &   &                    &                    \\
&$90$&$0.020$&$71.263\pm3.810$&$-19.860\pm36.642$&$-21.380\pm40.009$\\
&$89$&$0.043$&$77.258\pm2.966$&$-4.103\pm3.021$&$-4.170\pm3.312$\\
&$88$&$0.180$&$74.085\pm3.027$&$-0.502\pm4.471$&$0.033\pm4.160$\\
&$88$&$0.327$&$77.263\pm6.982$&$-3.547\pm15.760$&$-2.177\pm11.795$\\
&$89$&$0.493$&$82.143\pm8.881$&$-3.861\pm18.912$&$-1.664\pm10.408$\\
&$90$&$0.817$&$108.020\pm11.380$&$-0.372\pm5.025$&$0.306\pm2.510$\\
        & &   &                    &                    \\ \hline
\multirow{8}{5mm}{\begin{sideways}\parbox{25mm}{ 5 bins}\end{sideways}}
        & &   &                    &                    \\
&$108$&$0.021$&$70.236\pm2.907$&$-13.304\pm23.791$&$-13.990\pm25.630$\\
&$107$&$0.054$&$70.792\pm1.973$&$-0.528\pm1.393$&$-0.057\pm1.467$\\
&$107$&$0.254$&$81.099\pm3.568$&$-0.449\pm5.014$&$0.083\pm4.657$\\
&$106$&$0.438$&$93.850\pm9.979$&$-3.413\pm10.414$&$-2.427\pm8.840$\\
&$107$&$0.769$&$117.410\pm12.170$&$-1.407\pm3.107$&$-0.535\pm2.266$\\
        & &   &                    &                    \\ \hline
\multirow{8}{5mm}{\begin{sideways}\parbox{25mm}{~~~4 bins}\end{sideways}}
	& &    &			 &                  \\	
&$134$&$0.023$&$69.465\pm2.183$&$5.196\pm8.835$&$6.103\pm9.345$\\
&$132$&$0.098$&$71.411\pm1.535$&$-0.549\pm0.976$&$-0.052\pm0.979$\\
&$132$&$0.357$&$82.531\pm4.939$&$2.630\pm5.835$&$2.607\pm4.287$\\
&$132$&$0.703$&$109.330\pm8.647$&$-1.077\pm2.158$&$-0.269\pm1.541$\\
         & & &                    &                    \\ \hline
\multirow{7}{5mm}{\begin{sideways}\parbox{20mm}{~3 bins}\end{sideways}}
	   & & &			 &  \\
	   	   & & &			 &                   \\	
&$177$&$0.024$&$72.199\pm1.588$&$-1.432\pm1.553$&$-1.086\pm1.706$\\
&$176$&$0.216$&$75.981\pm1.886$&$-0.980\pm1.407$&$-0.389\pm1.274$\\
&$176$&$0.598$&$97.006\pm5.010$&$-0.785\pm1.288$&$-0.030\pm0.849$\\
          & & &                    &                    \\ \hline
\multirow{6}{5mm}{\begin{sideways}\parbox{14mm}{~~~~~2 bins}\end{sideways}}
	   & & &			 &                  \\	
&$264$&$0.026$&$71.913\pm1.105$&$-1.074\pm0.395$&$-0.684\pm0.418$\\
&$264$&$0.438$&$89.085\pm2.712$&$-0.895\pm0.496$&$-0.204\pm0.389$\\
          & & &                    &                    \\ \hline \hline
\end{tabular}
\label{tabresult}
\caption{Binned values for $H(z)$, $w(z)$ and $q(z)$, where $N$ and $z_{med}$ are respectively the number of data in each bin and
the median redshift in each bin.}
\end{table*}

\begin{figure*}
\centering
\includegraphics[width=0.9\textwidth]{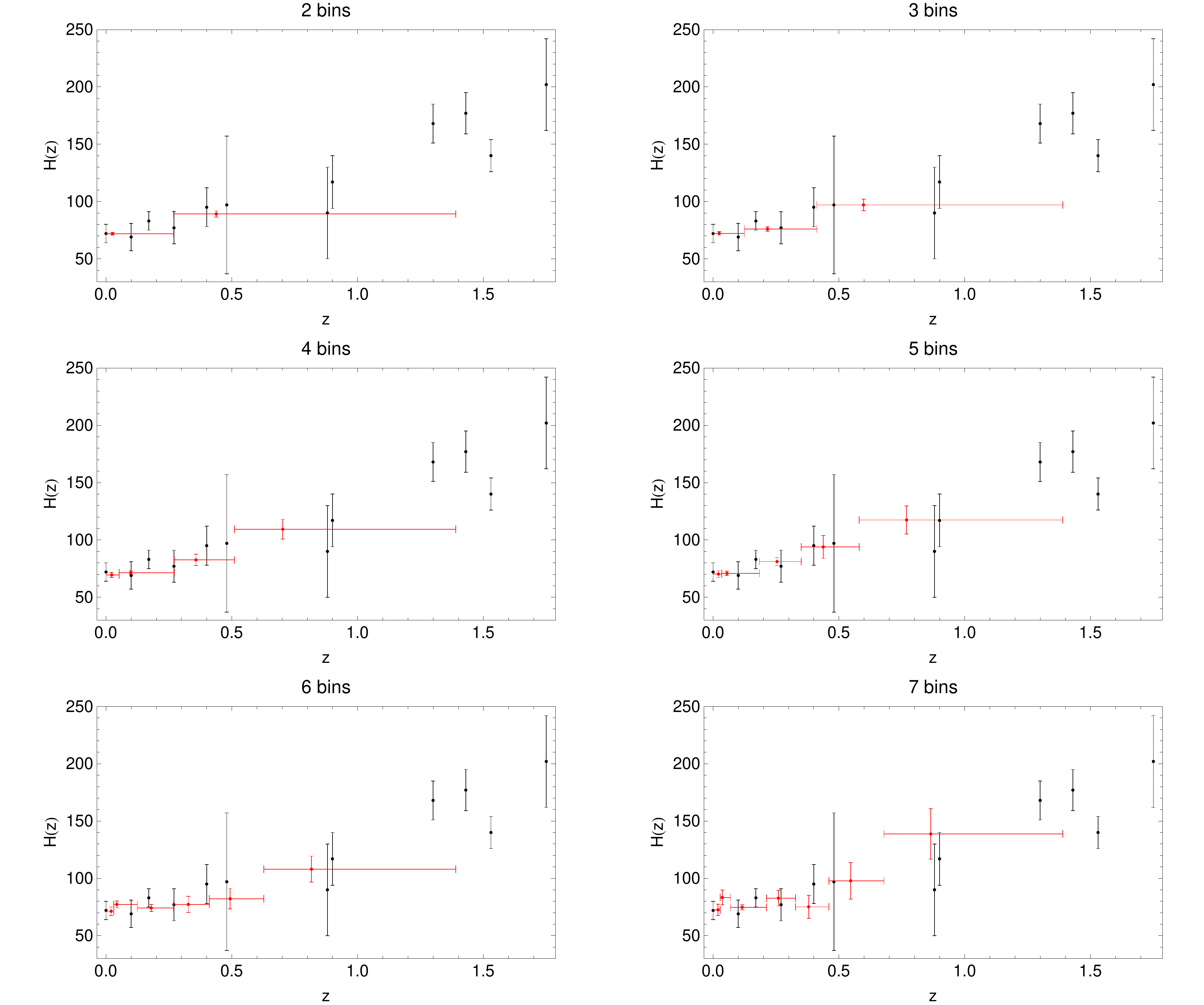}
\caption{Values of $H(z)$ for the binned approach for different values of independent bins, red points. The central point indicates the median
redshift of the bin with its corresponding error bars (vertical lines). The horizontal bars delimite the redshift bin. The $H(z)$ from \cite{Stern10} is also
plotted (black points) in order to compare the results}
\label{fig:Hbin}
\end{figure*}

\begin{figure*}
\centering
\includegraphics[width=0.9\textwidth]{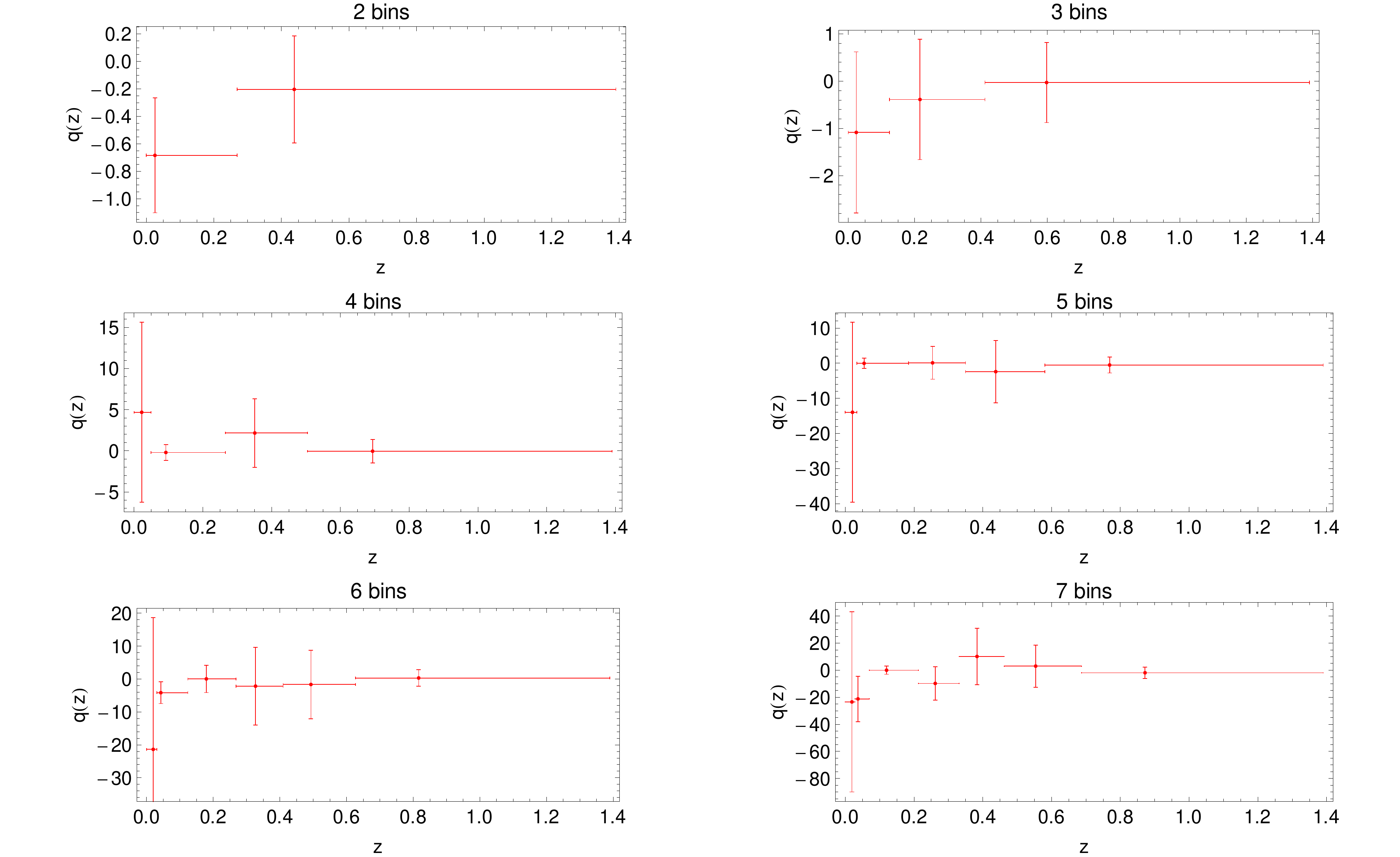}
\caption{Values of $q(z)$ for the binned approach for different values of independent bins, red points. The central point indicates the median
redshift of the bin with its corresponding error bars (vertical lines).}
\label{fig:qbin}
\end{figure*}

\begin{figure*}
\centering
\includegraphics[width=0.9\textwidth]{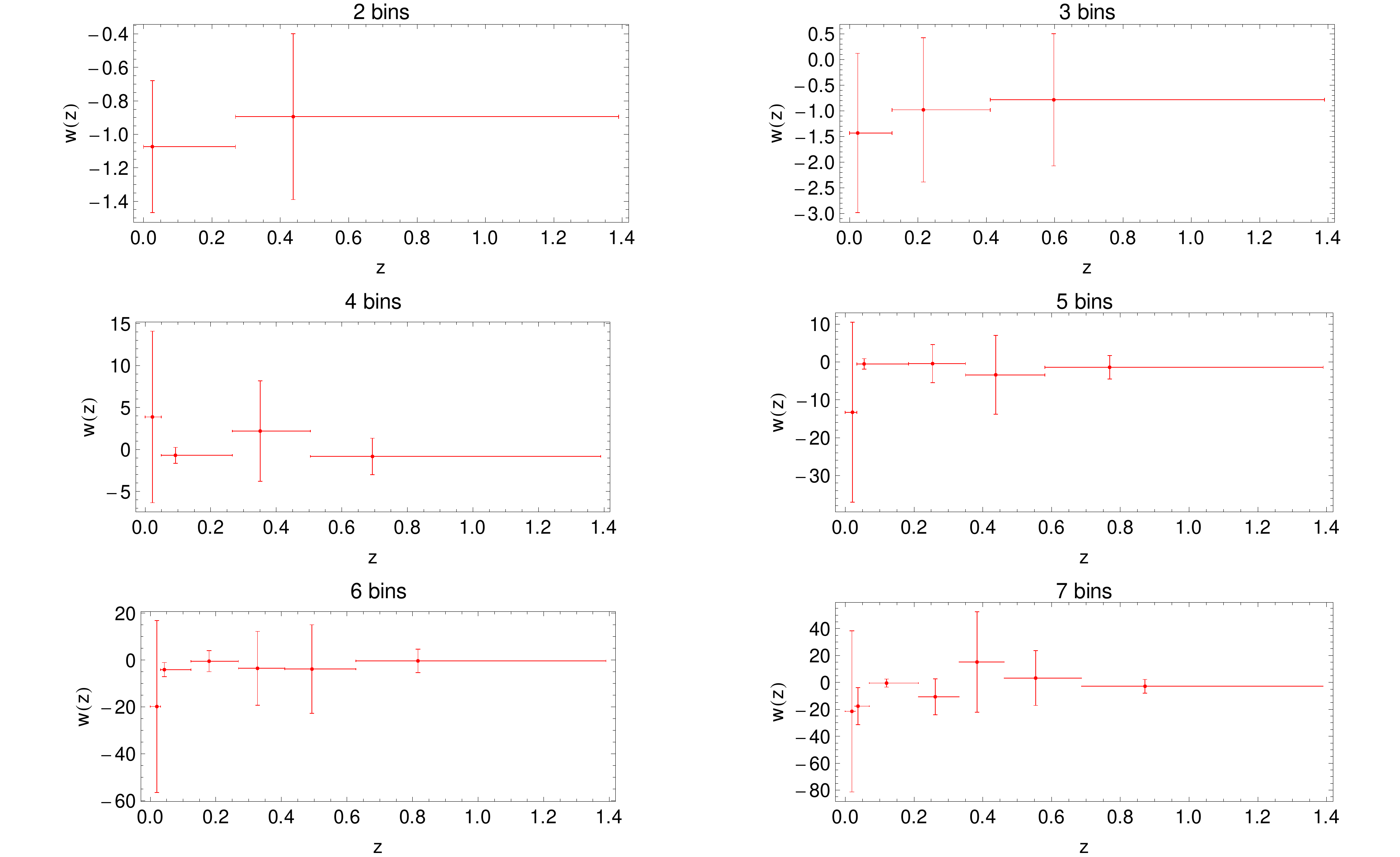}
\caption{Values of $w(z)$ for the binned approach for different values of independent bins, red points. The central point indicates the median
redshift of the bin with its corresponding error bars (vertical lines).}
\label{fig:wbin}
\end{figure*}

Results obtained with this approach are presented in Figs.~\ref{fig:Hbin}~-~\ref{fig:wbin} and in Table~\ref{tabresult}. For $H(z)$ the binned approach is well behaved, and the results agree with those presented in \cite{DalyDjorgovski3}. However, for the dark energy EoS parameter $w(z)$ and acceleration parameter $q(z)$,  we obtain large uncertainties when the
number of bins is large.  Nevertheless, these uncertainties get smaller as we decrease the number of independent bins and thus increase the amount data within a redshift bin.

\section{Summary and Conclusions}
\label{sec:Conclusions}

The abundant literature sources on model independent reconstructions of dark energy makes one believe this has become a
worthy line of investigation. The approach we have been concerned with in this paper has been amply covered
in \cite{DalyDjorgovski1,DalyDjorgovski2,DalyDjorgovski3}. The main idea is to take the dimensionless coordinate distance
and try to reconstruct it with minimal assumptions, and then push the reconstruction two levels deeper and
consider as well the first and second derivatives with respect to redshift. The way to do this is to carry out polynomial
fits in different redshift windows (with carefully chosen characteristics).

We have described the method following the original prescriptions, including throughout additional remarks to facilitate
comprehension of the various steps involved. Our main work tool has also been SNeIa luminosity measurements, but we have
been benefited by the availability of a sample which offers more than twice the number of sources: the very recent Union2 sample. This clearly allows
for some improvement, as increased density of the datasets induces in general tighter constraints. Nevertheless, it must be stressed that
the sample considered in  our paper is rather heterogeneous.

A fact that seems to have been overlooked in previous references is that the quite noticeable oscillations arising at
some levels of the reconstruction are mainly a consequence of data with large error bars in the samples. The fitting algorithm
does not seem to be the culprit for these effects. We have been able to reach these conclusions by considering
conveniently cut versions of the Union2 sample. The cut produces a spectacular reduction of the noisy features. This
suggests it could be worth revisiting this technique when significant advances occur in the compiled datasets.
For completeness and comparison we have also performed a similar analysis with GRBs, but the reconstruction turn out to
be rather poor (as otherwise expected).

On general grounds we can say the reconstruction from SNeIa is very satisfactory for the dimensionless coordinate, but it degrades
considerably as subsequent derivatives of it are considered. Estimates of global features agree very well with state of the art
guesses from other sources, and no strong dependence on the size of the window is appreciated at lowest levels.
We find a Hubble constant of a bit more than $70$ (km/s)/Mpc, a deceleration factor in the vicinity of $-0.5$, and a current
value of the dark energy parameter in the vicinity of $-1$.

For further insight into the problem we have also performed reconstructions using a mock SNeIa sample depicting a future optimistic
observational situation. The results support broadly what real data tell us, and in particular we find that dispersion in the data
influences   the reconstruction process a great deal.

And finally, using the binning method, we have obtained independent values of several quantities of interest. These may be useful
for a quick/preliminary estimate of cosmological constraints without having to resort to a full dataset.

\section*{Acknowledgements}
We are thankful to Diego Pav\'on for point out to us the interest of the works by Daly and Djorgovski.
Irene Sendra holds a PhD FPI fellowship contract
from the  Spanish Ministry of Economy and Competitiveness.
All three authors are supported
by the mentioned ministry
through research projects FIS2010-15492 and Consolider
EPI CSD2010-00064, and also by the Basque Government
through the special research action KATEA and the University of the Basque Country UPV/EHU under program UFI 11/55.


\begin{thebibliography}{1}%
\makeatletter
\providecommand \@ifxundefined [1]{%
 \ifx #1\undefined \expandafter \@firstoftwo
 \else \expandafter \@secondoftwo
\fi
}%
\providecommand \@ifnum [1]{%
 \ifnum #1\expandafter \@firstoftwo
 \else \expandafter \@secondoftwo
\fi
}%
\providecommand \enquote [1]{``#1''}%
\providecommand \bibnamefont  [1]{#1}%
\providecommand \bibfnamefont [1]{#1}%
\providecommand \citenamefont [1]{#1}%
\providecommand\href[0]{\@sanitize\@href}%
\providecommand\@href[1]{\endgroup\@@startlink{#1}\endgroup\@@href}%
\providecommand\@@href[1]{#1\@@endlink}%
\providecommand \@sanitize [0]{\begingroup\catcode`\&12\catcode`\#12\relax}%
\@ifxundefined \pdfoutput {\@firstoftwo}{%
 \@ifnum{\z@=\pdfoutput}{\@firstoftwo}{\@secondoftwo}%
}{%
 \providecommand\@@startlink[1]{\leavevmode\special{html:<a href="#1">}}%
 \providecommand\@@endlink[0]{\special{html:</a>}}%
}{%
 \providecommand\@@startlink[1]{%
  \leavevmode
  \pdfstartlink
   attr{/Border[0 0 1 ]/H/I/C[0 1 1]}%
   user{/Subtype/Link/A<</Type/Action/S/URI/URI(#1)>>}%
  \relax
 }%
 \providecommand\@@endlink[0]{\pdfendlink}%
}%
\providecommand \url  [0]{\begingroup\@sanitize \@url }%
\providecommand \@url [1]{\endgroup\@href {#1}{\urlprefix}}%
\providecommand \urlprefix [0]{URL }%
\providecommand \Eprint[0]{\href }%
\@ifxundefined \urlstyle {%
  \providecommand \doi [1]{doi:\discretionary{}{}{}#1}%
}{%
  \providecommand \doi [0]{doi:\discretionary{}{}{}\begingroup
  \urlstyle{rm}\Url }%
}%
\providecommand \doibase [0]{http://dx.doi.org/}%
\providecommand \Doi[1]{\href{\doibase#1}}%
\providecommand \bibAnnote [3]{%
  \BibitemShut{#1}%
  \begin{quotation}\noindent
    \textsc{Key:}\ #2\\\textsc{Annotation:}\ #3%
  \end{quotation}%
}%
\providecommand \bibAnnoteFile [2]{%
  \IfFileExists{#2}{\bibAnnote {#1} {#2} {\input{#2}}}{}%
}%
\providecommand \typeout [0]{\immediate \write \m@ne }%
\providecommand \selectlanguage [0]{\@gobble}%
\providecommand \bibinfo [0]{\@secondoftwo}%
\providecommand \bibfield [0]{\@secondoftwo}%
\providecommand \translation [1]{[#1]}%
\providecommand \BibitemOpen[0]{}%
\providecommand \bibitemStop [0]{}%
\providecommand \bibitemNoStop [0]{.\EOS\space}%
\providecommand \EOS [0]{\spacefactor3000\relax}%
\providecommand \BibitemShut [1]{\csname bibitem#1\endcsname}%
\bibitem{REVTEX41Control}%
  \BibitemOpen
  %
  \bibAnnoteFile{NoStop}{REVTEX41Control}%
\end{thebibliography}%


\begin{thebibliography}{99}

\bibitem{nobel}
  A.~G.~Riess {\it et al.} [ Supernova Search Team Collaboration ],
  Astron.\ J.\  {\bf 116 } (1998)  1009; 
  S.~Perlmutter {\it et al.} [ Supernova Cosmology Project Collaboration ],
  Astrophys.\ J.\  {\bf 517 } (1999)  565-586.

\bibitem{derev}
 D.~Sapone,
  Int.\ J.\ Mod.\ Phys.\  {\bf A25 } (2010)  5253;
  J.~Frieman, M.~Turner, D.~Huterer,
  Ann.\ Rev.\ Astron.\ Astrophys.\  {\bf 46 } (2008)  385;
  V.~Sahni, A.~Starobinsky,
  Int.\ J.\ Mod.\ Phys.\  {\bf D15 } (2006)  2105.


\bibitem{Model}
D. Huterer, G. Starkman, Phys. Rev. Lett. 90 (2003) 031301;
Y. Wang, M. Tegmark, Phys. Rev. Lett. 92 (2004) 241302;
A. Shafieloo, U. Alam, V. Sahni, A. Starobinsky, Month. Not. R. Astron. Soc. 366 (2006) 1081;
U. Alam, V. Sahni, A. Starobinsky, J. Cosmol. Astropart. Phys. 06 (2004) 008;
S. Nesseris, L. Perivolaropoulos, Phys. Rev. D 70 (2004) 043531;
D. Huterer, A. Cooray, Phys. Rev. D 71 (2005) 023506;
C. Shapiro, M.S. Turner, ApJ 649 (2006) 563;
M.S. Turner, D. Huterer, J. Phys. Soc. Japan 11 (2007) 111015;
U. Alam, V. Sahni, A. Starobinsky, J. Cosmol. Astropart. Phys. 11 (2007) 0702;
V. Sahni, A. Starobinsky, Int. J. Mod. Phys. D 15 (2006) 2105;
L. Perivolaropoulos, in AIP Conf. Ser. 848, Recent Advances in Astronomy and Astrophysics (New York: AIP) 698 (2006).

\bibitem{DalyDjorgovski1}
R.~A.~Daly, S.~G.~Djorgovski, Astrophys. J. 597 (2003) 9.

\bibitem{DalyDjorgovski2}
R.~A.~Daly, S.~G.~Djorgovski, Astrophys. J. 612 (2004) 652.

\bibitem{DalyDjorgovski3}
R.~A.~Daly, S.~G.~Djorgovski, Astrophys. J. 677 (2008) 1.

\bibitem{Schaefercrit}
  A.~C.~Collazzi, B.~E.~Schaefer, A.~Goldstein and R.~D.~Preece,
  arXiv:1112.4347 [astro-ph.HE],
  A.~C.~Collazzi, B.~E.~Schaefer and J.~A.~Moree,
  Astrophys.\ J.\  {\bf 729} (2011) 89.


\bibitem{GRB-No}
A. Diaferio, L. Ostorero, V. F. Cardone, arXiv:1103.5501

\bibitem{wcdm}
http://lambda.gsfc.nasa.gov/product/map/dr4/params \\
/wcdm\_sz\_lens\_wmap7.cfm

\bibitem{Amanullah10}
R.~Amanullah, {\it et al.}, Astrophys. J. 716 (2010) 712.

\bibitem{Kowalski08} M. Kowalski,
D.~Rubin, G.~Aldering, {\it et al.}, Astrophys. J. 686 (2008) 749.

\bibitem{Amanullah08}
R.~Amanullah, {\it et al.}, Astron. Astrophys. 486 (2008) 375.

\bibitem{Hicken09a}
M.~Hicken, {\it et al.}, Astrophys. J. 700 (2009) 331.

\bibitem{Holtzman08}
J~A.~Holtzman, {\it et al.}, Astron. J. 136 (2009) 2306.

\bibitem{Schaefer07}
B.~E.~Schaefer, Astrophys. J. 660 (2007) 16.

\bibitem{Riess:2011yx}
  A.~G.~Riess, L.~Macri, S.~Casertano, H.~Lampeitl, H.~C.~Ferguson, A.~V.~Filippenko, S.~W.~Jha and W.~Li {\it et al.},
  Astrophys.\ J.\  {\bf 730} (2011) 119
   [Erratum-ibid.\  {\bf 732} (2011) 129].

\bibitem{Cardone09}
V.~F.~Cardone, S.~Capozziello, M.~G.~Dainotti, Month. Not. R. Astron. Soc. 440 (2009) 775.

\bibitem{Stern10}
D. Stern, R. Jimenez, L. Verde, M. Kamionkowski, A. Stanford,  JCAP  1002 (2010) 008.

\bibitem{Wang:2008zh}
Y. Wang, Phys. Rev. D 77 (2008) 123525.

\bibitem{Albrecht06}
A. Albrecht, (2006) astro-ph/0609591.

\bibitem{Chevallier01}
M. Chevallier, D. Polarski, David, Int. J. Mod. Phys. D 10 (2001) 213.

\bibitem{Linder03}
E. V. Linder, Phys. Rev. Lett. 90 (2003) 091301.

\bibitem{Escamilla11}
C.~Escamilla-Rivera, R.~Lazkoz, V.~Salzano and I.~Sendra,
  JCAP {\bf 1109} (2011) 003.

\bibitem{wfirst}
  J.~Green, P.~Schechter, C.~Baltay, R.~Bean, D.~Bennett, R.~Brown, C.~Conselice and M.~Donahue {\it et al.},
  arXiv:1108.1374 [astro-ph.IM].

\bibitem{JDEM}
A.~Albrecht, L.~Amendola, G.~Bernstein, D.~Clowe, D.~Eisenstein, L.~Guzzo, C.~Hirata, D.~Huterer {\it et al.},
[arXiv:0901.0721]

\bibitem{SNfactory} G. Aldering, et al., {http://snfactory.lbl.gov/snf/pdf/} \\ spie$\_$2002.pdf

\bibitem{snerror}
A.G. Kim, E.V. Linder, R. Miquel, N. Mostek, Month. Not. R. Astron. Soc. 347 (2004) 909;
A. Upadhye, M. Ishak, P. Steinhardt, Phys. Rev. D 72 (2005) 063501;
M. Ishak, Found. Phys. 37 (2007) 1470.

\bibitem{Sendra:2011pt}
  I.~Sendra, R.~Lazkoz,
  arXiv:1105.4943 [astro-ph.CO] (MNRAS accepted).


\end{thebibliography}
\end{document}